\documentclass[a4paper,fleqn]{cas-dc}

\usepackage[authoryear,longnamesfirst]{natbib}
\usepackage{caption}
\usepackage{multirow}
\usepackage{makecell, cellspace, caption}
\usepackage{balance}
\def\tsc#1{\csdef{#1}{\textsc{\lowercase{#1}}\xspace}}
\tsc{WGM}
\tsc{QE}
\tsc{EP}
\tsc{PMS}
\tsc{BEC}
\tsc{DE}

\begin{document}
\let\WriteBookmarks\relax
\def\floatpagepagefraction{1}
\def\textpagefraction{.001}

\shorttitle{}

\shortauthors{Bibekananda Nath et~al.}

\title [mode = title]{Broadband High-Temperature Multilayer Pyramid-Shaped Metamaterial Thermal Absorber for Thermophotovoltaic applications}                      



%
\author[1]{Bibekananda Nath}[auid=000,bioid=1,
                        orcid=0000-0001-7511-2910]





\credit{Conceptualization, Methodology, Visualization, Software, Investigation, Writing – original draft}

\affiliation[1]{organization={Bangladesh University of Engineering and Technology},
    addressline={}, 
    city={Dhaka},
    postcode={1205}, 
    country={Bangladesh}}

\author[1]{Ahmed Zubair}[orcid=0000-0002-1833-2244]
\cormark[1]
\credit{Conceptualization, Methodology, Visualization, Resources, Writing – original draft, Writing – review \& editing, Supervision}
\ead{ahmedzubair@eee.buet.ac.bd}
\ead[URL]{http://ahmedzubair.buet.ac.bd/}





\cortext[cor1]{Corresponding author}



\begin{abstract}
A broadband, thermally stable absorber is essential for thermophotovoltaic (TPV) systems to simultaneously convert solar and industrial waste heat into usable energy to meet growing power demands. Here, we proposed an ingenious polarization-independent truncated pyramid-shaped symmetric multilayer metamaterial absorber in a metal-insulator-metal-insulator (MIMI) architecture with almost complete absorption over a broad wavelength range. A total of six structures (W/AlN, Mo/AlN, Ta/AlN, Rh/MgO, Rh/SiO\textsubscript{2}, Re/BN) were designed, and the materials were selected based on their lattice matching to prevent delamination at interfaces between layers. The absorption mechanism was studied at room temperature using the finite difference time domain (FDTD) method, and the structure was optimized through a brute force design approach, which illustrates a best average absorption of 98.2\% till 4000 nm and 97.73\% till 5072 nm wavelength for the W/AlN structure with metal and dielectric thicknesses of 60 nm and 17.5 nm, respectively. Moreover, W/AlN structure exhibits over 96\% average absorption up to 50$^\circ$ incident angles irrespective of polarizations. The thermal stability was evaluated using the finite element method (FEM) by determining von Mises stress at elevated temperatures. Thermal analysis revealed that only W/AlN can withstand around 1700 K temperature and 1500 times the incident power before permanent deformation. A temperature-dependent Drude-Lorentz model was used to further analyze the effect of absorption on the optical performance of the highly absorptive and thermally stable W/AlN structure. Additionally, we determined the effect of the concentration factor, and the operating temperature on the system efficiency by considering the emission loss of the heated absorber. 
This research has enormous potential in high-temperature applications like thermal energy storage systems, photodetectors, and sensors.
\end{abstract}


\maketitle

\section{Introduction}

Thermophotovoltaic (TPV) systems harness energy from heat sources, converting the radiated heat flux from the sun and the absorbed heat through conduction, convection, or radiation processes from the industrial furnace to electricity through photovoltaic cells. For efficient conversion from both heat sources simultaneously, a broadband absorber is required that needs to maintain a constant high temperature, as this will act as a source of photons for the photovoltaic cell followed by a narrowband emitter. 
However, one of the most significant issues in such cases is absorbing low-energy photons from industrial waste heat.  As a large number of industries generate waste heat at a temperature range of about 400 to 2000 K, their emitted radiation is made up of photons that fall beyond the visible spectrum \cite{thekdi2015industrial}. To utilize these photons, an absorber is required, which must absorb radiation over a wide range of wavelengths. Moreover, the absorber must have the capability to withstand high temperatures to develop a highly efficient TPV system \cite{rephaeli2009absorber}. Depending on the spectral bandwidth of the incident light, two types of absorbers are possible based on their applications: narrowband and wideband absorbers \cite{azri2023advancement}. Generally, narrowband absorbers are used in visible to infrared range sensing applications \cite{luo2016perfect, cheng2021plasmonic, yu2016nonlinear, sun2019wafer, wu2011large}. On the other hand, the broadband absorbers are better for energy-generating applications such as photodetectors and solar cells \cite{chen2023ultra, zhou2021nanobowls, Ishfak21OMEx, Partha23NA}, solar thermal applications \cite{rana2023broadband, zhou2020ultra, song2022ultra, huang2022ultra, Nabila22AA, Nowshin23OptCon, wan2015broadband, wang2021broadband}.

\begin{figure*}
	\centering
		\includegraphics[scale=.53]{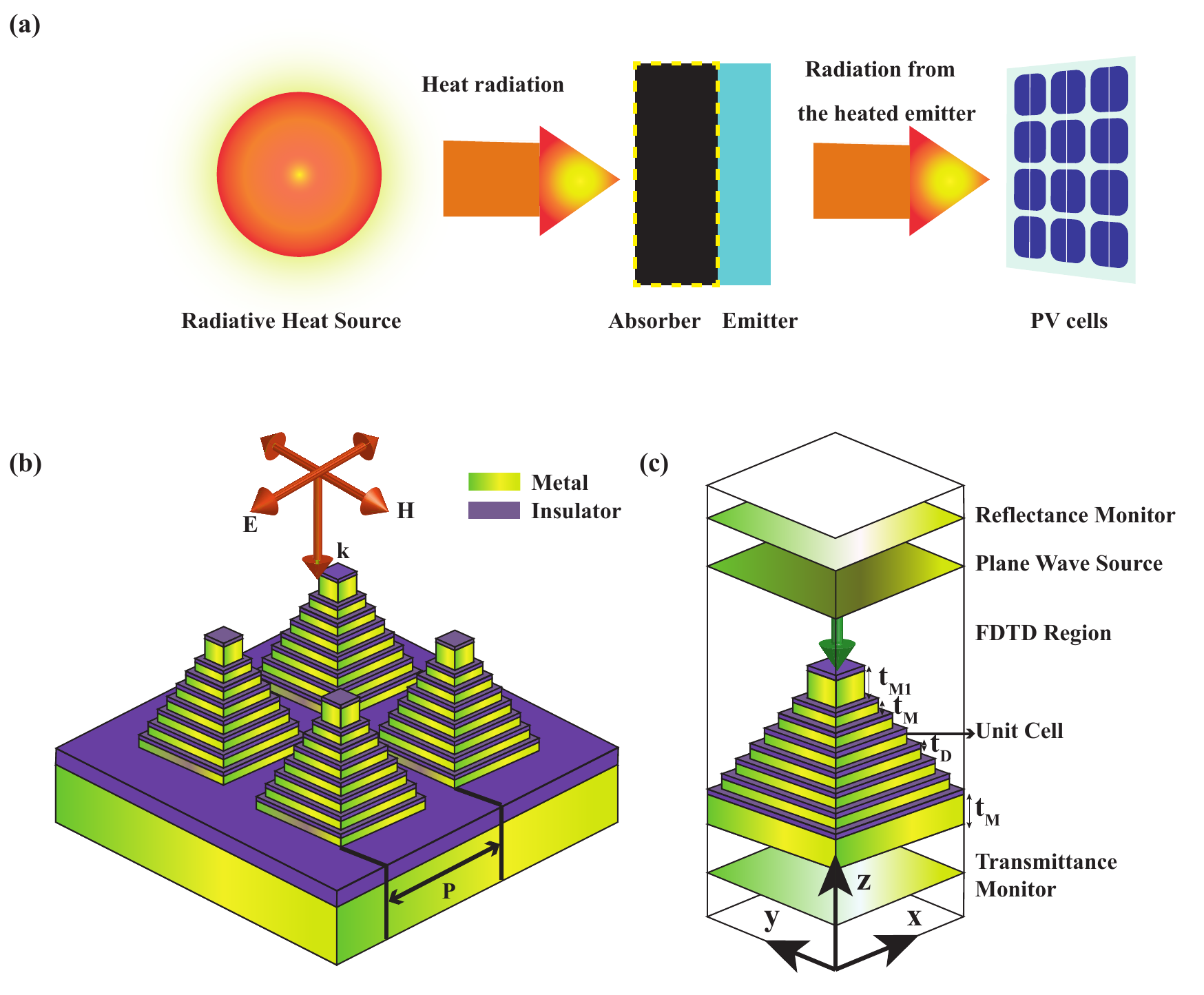}
	\caption{ Schematic diagram of the (a)  overall TPV system with an absorber-emitter structure, (b) periodic structure of the proposed metamaterial absorber, and (c) unit cell of the proposed absorber structure inside the simulation setup.}
	\label{FIG:1}
\end{figure*}
To utilize the industrial waste heat, several research studies were performed in the past years \cite{utlu2013investigation,utlu2020thermophotovoltaic}. Most of them are associated with an emitter and filter section for spectral control between the source and the cell to utilize only the conductive or convective heat energy and the overall TPV system used for industrial waste heat recovery was either flat-type \citep{li2022design} or stereoscopic-type \citep{saraey2022evaluation}. However, a vast amount of heat flux is radiated through high-temperature (typically 600–1500 $^\circ$ C) exhaust gases from the heating furnaces. Even boilers and heaters of several coal-fired power plants of various capacities can produce a large amount of waste heat with a higher percentage of heat loss at around 200-550 $^\circ$ C \citep{rashid2020recent}. Hence, an efficient absorber is essential to exploit the radiated heat energy of both the sun and the exhaust gas of industrial furnaces.  Fig.\,\ref{FIG:1}(a)  represents a schematic diagram of the overall TPV system  containing the absorber-emitter structure. The absorber is heated from a concentrated radiative heat source up to a certain temperature.  The narrowband emitter is thermally connected to the absorber to maintain the same temperature as the absorber and is spectrally matched with the PV cell. The heated emitter re-radiates photons towards the PV cells.

Several studies have been conducted on developing a broadband absorber with high average absorption using photonic crystals \cite{luo2019near, kang2018wideband}, gratings \cite{wu2016polarization, zhang2022grating}, metasurfaces \cite{Sarkar2024MA},  nanowires \cite{wang2022ultra}, quasi-periodic nanocones \cite{huo2018broadband}, nanoholes \cite{zou2024ultra}, plasmonic nanoparticles \cite{Nowshin23OptCon, cai2019anisotropic, perdana2022thin,Nowshin23RINP}. Among them, metamaterial structures exhibit overall better average absorption because of their unique ability to enhance the light-matter interaction and create different resonance phenomena. Due to the diffraction limit, this plasmonic behaviour can only be obtained in nanoscale devices like metamaterials \cite{Nowshin23OptCon, Sarker:21}. The light-matter interaction continues to drive advancements in energy and environmental applications. In recent years, there has been a surge of research on nanophotonic and metamaterial absorbers tailored for real-world applications. Ngobeh \textit{et al.} demonstrated a multilayer metamaterial structure with high average absorption from UV to around 2500 nm wavelength range, designed for solar thermal applications \cite{ngobeh2025numerical}. Schmitz \textit{et al.} used carbon-based electrodes derived from biowaste and a n-butylammonium cap layer to create a 2D/3D interface with the Cs\textsubscript{2}AgBiBr\textsubscript{6} perovskite solar cell to improve hole extraction and band alignment \cite{schmitz2024improved}. The solar cell developed by them was highly efficient without any hole transport material (HTM). These works highlight how engineered light–matter interactions now underpin scalable, multifunctional devices that address both energy sustainability and environmental resilience.

For efficient light absorption, several studies have been conducted to find a perfect absorber based on metal-dielectric structures. A near-perfect absorber having an average absorption of more than 94\% was reported by Sayed \textit{et al.} \citep{sayed2023design}. Their proposed MIM structure was made of Mn-SiO\textsubscript{2}-Mn based on particle swarm optimization analysis, which could only absorb visible range light. To improve absorption above the visible region, a Cr-based metasurface absorber for the thermophotovoltaic system was proposed by Rana \textit{et al.}, which exhibits an average absorption of around 90\% for (300-1200)nm \citep{rana2023broadband}. A huge improvement in the search for wide-band perfect absorption was reported by Soliman \textit{et al.}\,\citep{soliman2023broadband}. 
They proposed Ni-SiO\textsubscript{2}-Ni-SiO\textsubscript{2}, MIMI metasurface absorber and obtained an average absorption of 99\% for (200-2000) nm. To improve absorption via structural anisotropy, a broadband absorber based on hyperbolic metamaterials was designed by Qin \textit{et al.} made of TiN-SiO\textsubscript{2} nanopillar arrays over a Ti film that produced an average absorption of over 90\% within (300-2000) nm \citep{qin2020ultra}. Similarly, Li \textit{et al.} designed a nanopillar structure made of TiN-SiO\textsubscript{2} and obtained an average absorption of over 97\% from (300-2100) nm \citep{li2023theoretical}. Researchers worked on multilayer metamaterial structures to extend absorption over a wide range of wavelengths. These structures can absorb photons of different wavelengths at different layers via SPP. Hoa \textit{et al.} proposed a perfect absorber based on a multilayer Au-Si disc, which was capable of absorbing 90\% of the incident spectra over (480-1480)nm \cite{8598923}. Qin \textit{et al.} designed a multilayer metamaterial based on Al\textsubscript{2}O\textsubscript{3}/W, IMIM absorber that provided an average absorption of over 90\% for (420-2112) nm \citep{qin2022broadband}. Cheng \textit{et al.} demonstrated how absorption can be improved by placing ring-shaped W arrays over SiO\textsubscript{2}-W structure through plasmon resonance and Fabry-Perot resonance and reported an average absorption of more than 90\% within (300-2000) nm \citep{cheng2021ultra}. However, these MIM structures cannot provide high absorption at shorter wavelengths. To overcome this issue, Li \textit{et al.} formed a composite metamaterial structure by introducing an IMI grating over MIM structure, which has an average absorption of around 90\% up to 2800 nm for normal incidence and obtained an average of total absorption of over 80\% for up to 50$^\circ$ oblique incidence \cite{li2019ultra}. However, the low average absorption of IMIM composite structure was eliminated by Soliman \textit{et al.}\,\cite{soliman2023broadband}. They reported a thermophotovoltaic metamaterial absorber made of SiO\textsubscript{2}-Ni-SiO\textsubscript{2}-Ni (IMIM) structure and obtained an average efficiency of 99.18\% within (250-2000) nm wavelength for the TE and TM modes and an oblique angle of up to 60$^\circ$. 
 
However, most of these works focused only on the perfect absorption of visible to near-IR radiation and did not consider the thermal performance of their proposed structure; however, at elevated temperatures, nanostructures are subjected to high thermal strain if the metal and dielectric materials are not phase-matched. In such a case, the structure is more likely to be cracked \cite{dias2023photonics}. Apart from this, high temperatures will cause higher thermal stress at the interface between the two different materials, which may cause the materials to melt and possibly cause chemical reactions at a much lower temperature than the melting point, which leads to the delamination of the layered structures.

To simultaneously utilize both industrial waste heat along with solar radiation, we propose a total of six novel multilayer pyramid-shaped broadband metamaterial TPV absorbers based on metals (W, Mo, Ta, Rh, Re) and dielectrics (AlN, SiO\textsubscript{2}, BN, MgO). Materials with high melting point temperatures and high thermal conductivity, along with almost similar thermal expansion coefficients and lattice constants, were chosen to avoid interfacial deformation at high temperature. The optical and thermal performances of the designed structures were assessed via the FDTD method by solving Maxwell's equations in 3D space and by comparing the material's ultimate tensile strength with the von Mises thermal stress obtained through the FEM analysis, respectively. The temperature-dependent optical performance of the best-performing structure based on optical and thermal analysis was further determined by using a temperature-dependent Drude-Lorentz model. In addition to harnessing energy from waste heat and the sun, the proposed metamaterial absorber study can be applied to hyperspectral imaging, sensors, and photodetectors.

\section{Structure design and simulation methodology}
\begin{figure*} [t]
	\centering
		\includegraphics[scale=.53]{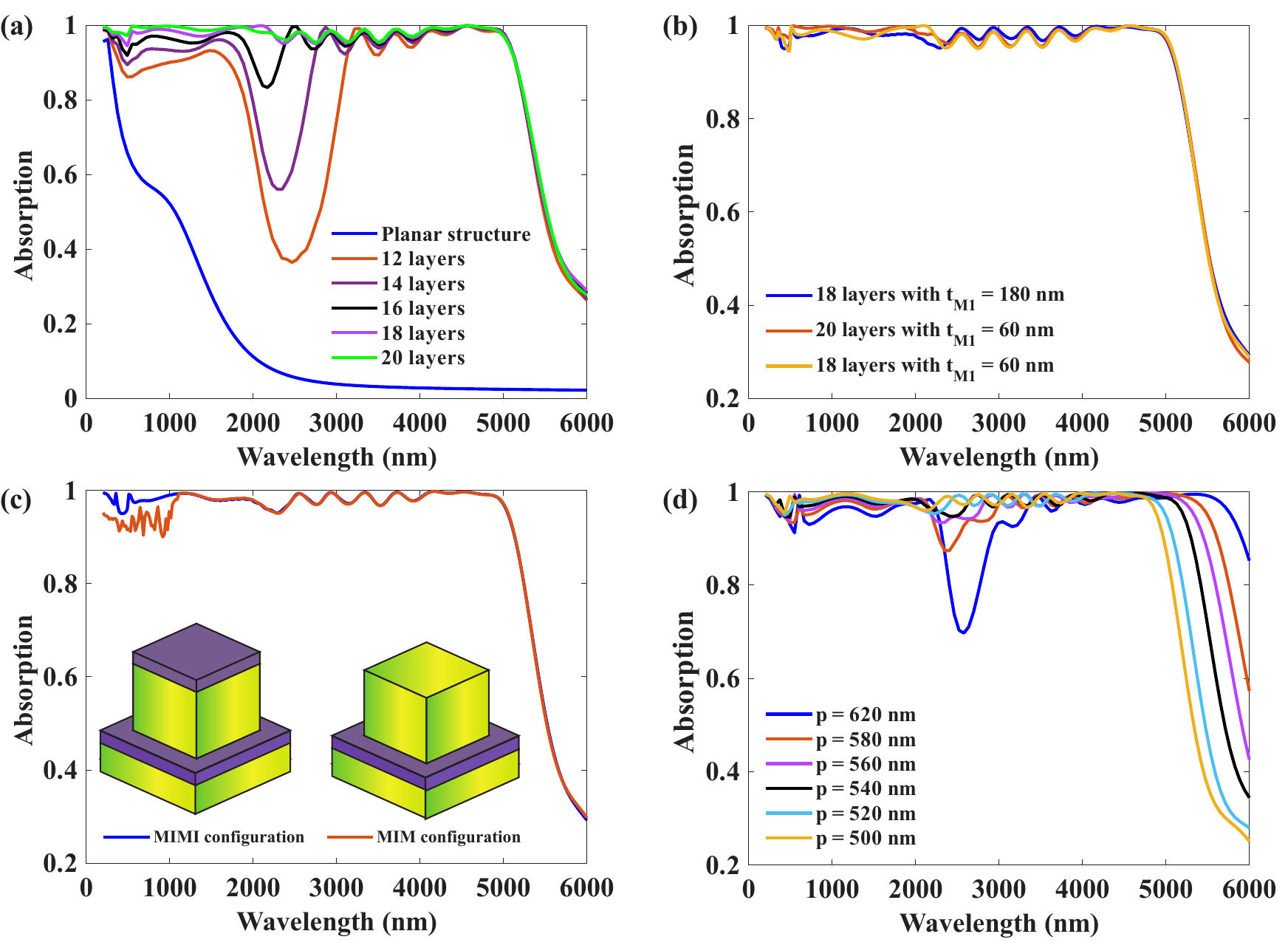}
	\caption{The performance of the proposed metamaterial absorber for (a) various numbers of layers, (b) various thicknesses of the top metal layer on absorption, (c) MIMI and MIM structures, and (d) different periodicity of the proposed structure.}
	\label{FIG:2}
\end{figure*}
We designed a truncated pyramid-shaped structure consisting of vertically stacked square slabs made of metal and dielectric materials. Fig.\,\ref{FIG:1}(b) illustrates the schematic diagram of the proposed absorber. Fig.\,\ref{FIG:1}(c) represents the setup of FDTD simulation of a unit cell of the proposed structure. The structure substrate was designed with a thick metal layer with thickness t\textsubscript{M1} and an ultrathin dielectric layer with thickness t\textsubscript{D} on top of it. Other metal layers were of identical thickness t\textsubscript{M2}. The thick metallic substrate acts as an opaque layer to reduce transmission and improve absorption, while the dielectric layer improves the magnetic coupling of the structure, causing a slow-wave effect. Above the substrate, alternate layers of metal and dielectric were placed so that each metal and its upper dielectric layer have an identical width; however, the widths of subsequent metal-dielectric pairs decrease as the number of layers increases. Although all of the metal layers were kept at the same thickness, the thickness of the topmost metal layer was three times the usual thickness to improve the overall absorption by localized surface plasmon resonance (LSPR) for an extended period. All structural parameters, such as the thickness and width of the metal and dielectric layers, the number of composite layers, and the periodicity of the structure, were optimized using the brute force parameter sweep approach. In this study, tungsten (W), tantalum (Ta), rhenium (Re), rhodium (Rh), and molybdenum (Mo) were chosen as the metals because of their outstanding high-temperature properties (see Table S2 of Supplementary Material).
This property is essential to maintaining the performance at varying temperatures. We chose boron nitride (BN) as the dielectric material, Re was used as the metal, and aluminum nitride (AlN) was chosen as the dielectric material for W, Ta, and Mo. On the other hand, MgO and SiO\textsubscript{2} were chosen as dielectrics for Rh based on their lattice parameter matching to avoid larger mismatches in thermal properties, which can lead to interfacial deformation and cracking at high temperatures due to high strain and excessive thermal stress. The lattice parameters of the used materials in this study are provided in Table S1 of the Supplementary Material.

Optical performance was analyzed by solving the Maxwell equation using the FDTD method, which required complex refractive index data. The complex refractive index of W and SiO\textsubscript{2} was taken from Palik \textit{et al.} \cite{palik1998handbook}. The extinction coefficient and refractive index of Mo and Ta were taken from Ordal \textit{et al.} \cite{ordal1988optical}. For Rh and Re, those data were adopted from Windt \textit{et al.} \cite{windt1988optical}. The same data for AlN and MgO were taken from Kischkat \textit{et al.} \cite{kischkat2012mid} and Stephens \textit{et al.} \cite{stephens1952index}, respectively. The complex refractive index data for BN was collected from Lee \textit{et al.} \cite{lee2019refractive}, and were extrapolated for simulation (the complex refractive index vs wavelength plot for all metals and dielectrics used in this study have been provided in Figs. S1 and S2 of the Supplementary Material). The complex refractive the 3D-FDTD simulation was performed in the FDTD module of the Ansys Lumerical software. Because of the periodicity, the whole simulation was conducted over a unit cell. In the simulation interface, the periodic boundary condition was applied to incorporate the structure periodicity along the x and y directions. The plane wave source was placed along the z-direction so that the incident radiation from the source was in the downward direction. To minimize the reflection from the illuminated direction and to absorb most of the transmitted and reflected wave from the simulation environment, 64 perfectly matched layers (PML) were used along the z-direction. The setup of the Lumerical simulation environment is illustrated in Fig.\,\ref{FIG:1}(c). 

A conformal mesh was used to refine the simulation accuracy because of the presence of metal. The simulation was conducted for transverse electric (TE), transverse magnetic (TM), and unpolarized plane wave sources. To evaluate the perfect broadband absorption, the wavelength of the source radiation spectrum was chosen from 200 to 6000 nm. For accurate and time-effective simulation, mesh sizes of 2.5 nm along the z direction and 5 nm along both the x and y directions were defined. To enhance the resolution of the simulation interface by minimizing the numerical dispersion, five mesh cells were created per wavelength for the surrounding region. Frequency domain power monitors were employed to enumerate reflected and transmitted waves from the designed metamaterial structure, as depicted in Fig.\,\ref{FIG:1}(c). Field power along the spatial directions was calculated in the power monitors wavelength ($\lambda$) domain. Two separate monitors were used for the calculation of reflected and transmitted waves according to the following equations:
\begin{equation}
    R^{\prime}(\lambda) = \frac{P^{\prime}_R}{P_I}, 
\end{equation}
\begin{equation}
    T^{\prime}(\lambda) = \frac{P^{\prime}_T}{P_I}.
\end{equation}
The transmittance and reflectance from the designed structure are functions of wavelength and are represented by $R^{\prime}(\lambda)$ and $T^{\prime}(\lambda)$, respectively. Besides, $P_I$, $P^{\prime}_R$, and $P^{\prime}_T$ indicate the incident, reflected, and transmitted wave's power, respectively. The absorption depends on the reflected and transmitted waves, and it is also wavelength-dependent. The absorption ($A^{\prime}(\lambda)$), was determined as follows:
\begin{equation}
    A^{\prime}(\lambda) = 1-(R^{\prime}(\lambda) + T^{\prime}(\lambda)),
\end{equation}

 The thermal performance of the designed structure was evaluated in the COMSOL Multiphysics software using the FEM technique by solving the heat transfer equation in 3D space. The model used an external radiation source with irradiance for multiple bands to account for the effect of various heat sources of different temperatures, including the sun. The impact of high temperature was measured based on the von Mises thermal stress on the structure.
For determining the thermal stress, a steady-state analysis was performed by solving the Lagrangian form of Newton's second law as follows \citep{bonet1997nonlinear},
\begin{equation}
  \rho  \frac{\partial^2 u}{\partial t^2} = F_V + \nabla_X [(I+\nabla u)S]
  \label{{eq:heat}}
\end{equation}
\\\
\begin{figure*}[h]
	\centering
		\includegraphics[scale=.53]{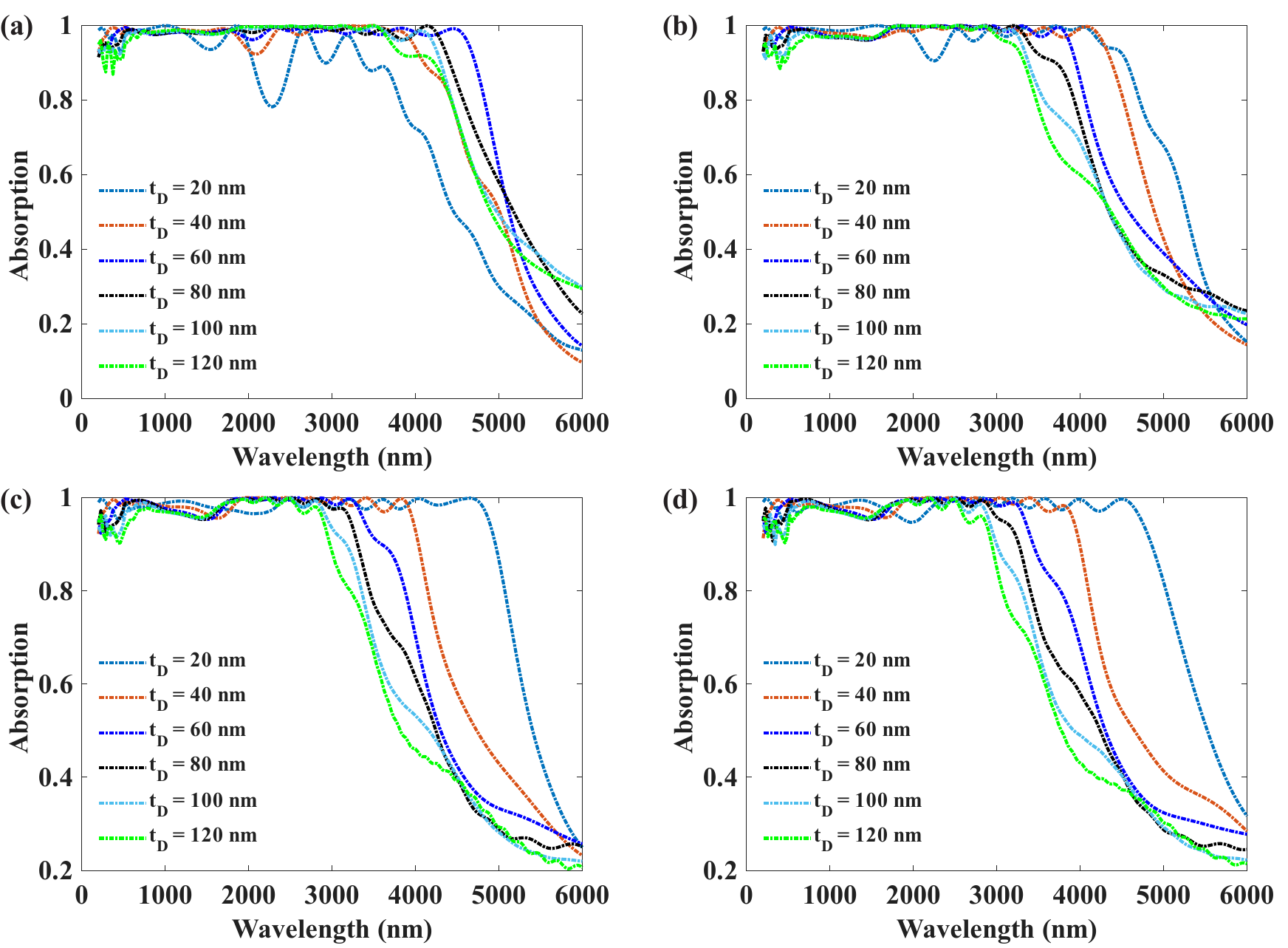}
	\caption{absorption profile for different dielectric thickness with metal thicknesses of (a) 20 nm, (b) 40 nm, (c) 60 nm, and (d) 70 nm. }
	\label{FIG:3}
\end{figure*}
\begin{figure*}[h]
	\centering		\includegraphics[scale=.53]{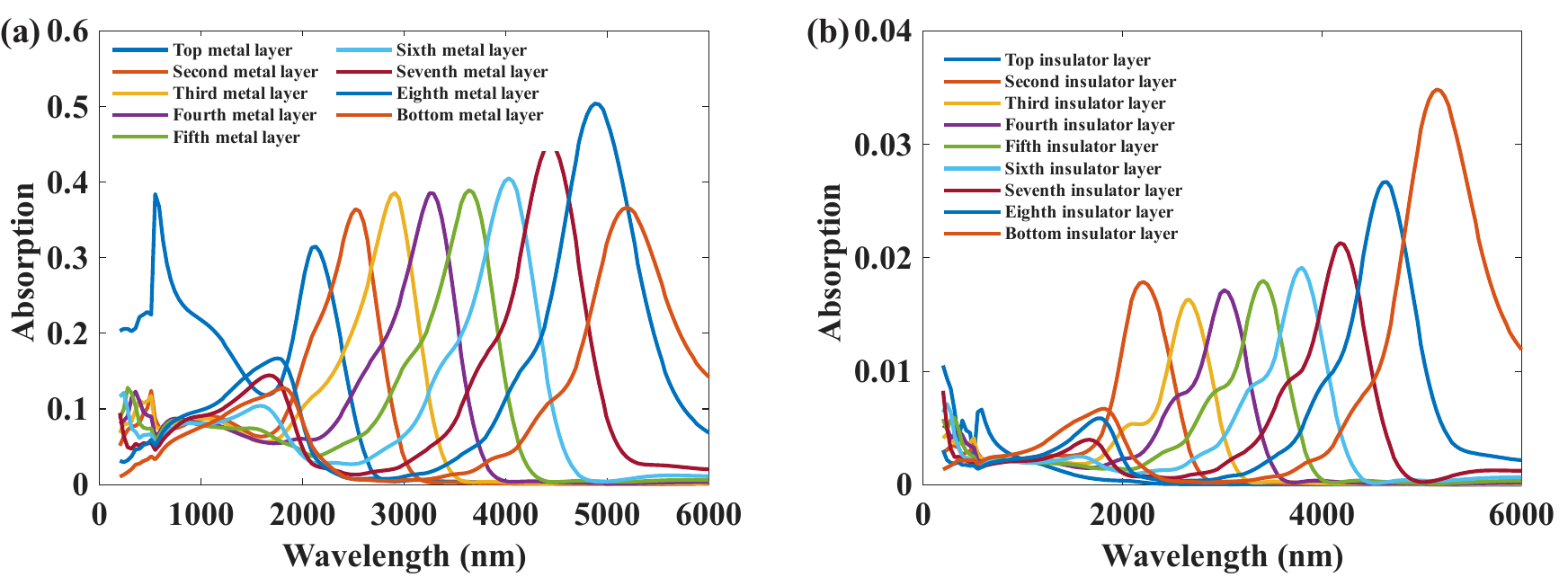}	\caption{absorption for (a) different metal and (b) different insulator layers, indicating the presence of plasmon resonance at different metal-insulator junctions, because of which the resonance peak shifts towards the longer wavelength for the bottom layers.}
	\label{FIG:4}
\end{figure*}
\begin{figure}
	\centering
		\includegraphics[scale=.58]{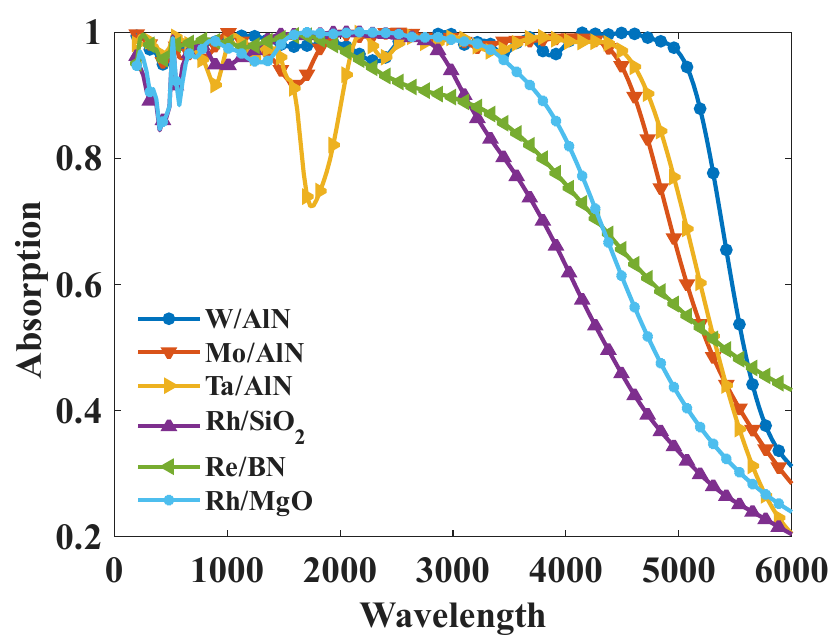}
	\caption{absorption profile of different meal-dielectric composites}
\label{FIG:5}
\end{figure}
During the thermal stability computation of the proposed composite structure, alternate layers of metal and dielectric with close lattice matching were strongly connected to support the multilayer structure from delamination. These fixed contacts can apply compressive force against the natural expansion of the material due to increasing temperature and can create thermal stress at the interface between two layers. This stress depends on the deformation gradient of the material according to the second Piola-Kirchhoff stress factor \citep{bonet1997nonlinear}, 
\begin{equation}
    S = J \sigma F^{-1} F^{\tau}
    \label{eq5}.
\end{equation}

Here, $\tau= J \sigma$, is the Kirchhoff stress tensor. \textit{J }is the volume factor, which indicates the change in volume of the structure due to deformation. \textit{F} is the tensor of the deformation gradient. If the value of this compressive stress exceeds the maximum tensile strength of the material, the material will begin to deform permanently from its actual shape and a fracture will occur. 
\begin{table*}[width=2\linewidth,cols=4]
\caption{Average absorption of different metal-dielectric structures at various wavelength regions.}\label{tbl1}
\begin{tabular*}{\tblwidth}{@{} LLLLL@{} }
\toprule
\makecell {Structures} & \makecell{UV Region\\(200 -- 400 nm)} & \makecell{Visible Region \\ (400 -- 780 nm)} & \makecell{ Near IR Region \\ (780 -- 2500 nm)} & \makecell{Broadband \\ (200  -- 5072 nm}\\
\midrule
\makecell{W/AlN} & \makecell{95.1\%} & \makecell{97.38\%} & \makecell{98.25\%} & \makecell{97.73\%}\\
\makecell{Mo/AlN} & \makecell{93.64\%} & \makecell{97.57\%} & \makecell{97.82\%} & \makecell{87.52\%}\\ 
\makecell{Ta/AlN} & \makecell{92.56\%} & \makecell{98.26\%} & \makecell{91.46\%} & \makecell{86.54\%}\\
\makecell{Rh/MgO} 
& \makecell{91.5\%} & \makecell{95.05\%} 
& \makecell{98.36\%} & \makecell{76.62\%}\\ 
\makecell{Rh/SiO\textsubscript{2}} & \makecell{93.75\%} & \makecell{93.67\%} & \makecell{98.85\%} &\makecell{68.9\%} \\
\makecell{Re/BN} & \makecell{93.11\%} & \makecell{97.81\%} & \makecell{90.11\%} &\makecell{79.75\%} \\
\bottomrule
\end{tabular*}
\label{tab:1}
\end{table*}
\begin{figure*} [h]
	\centering
		\includegraphics[scale=.53]{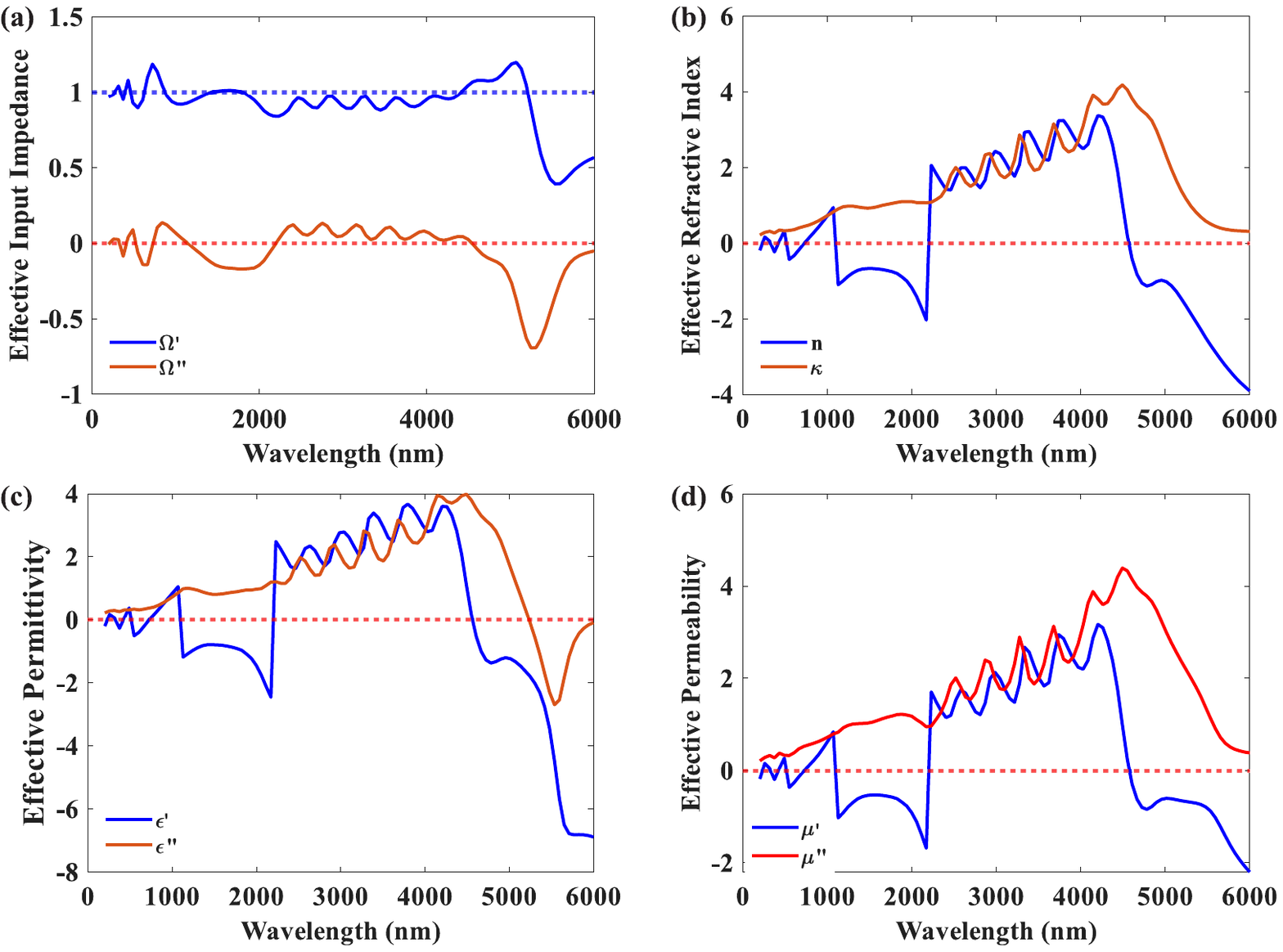}
	\caption{Proposed absorber's real and imaginary value of (a) effective input impedance, (b) effective refractive index, (c) effective permittivity, and (d) effective permeability.}
	\label{FIG:6}
\end{figure*}
Here, the von Mises stress was calculated by solving the momentum-balance equation for the elastic wave inside the structure. Considering this, the Lagrangian form of Newton's second law for our structure's stress analysis can be written in the following form \citep{bonet1997nonlinear}:
\begin{equation}
\rho  \frac{\partial^2 u^{\prime}}{\partial t^2} = F_V^{\prime} + \nabla_X P
  \label{lagragianformnewton}
\end{equation}
In the above equation, $P = (I^{\prime}+\nabla^{\prime})S$, the first Piola-Kirchhoff stress tensor of our system, where $\nabla ^{\prime}$ is the displacement gradient and ${I^{\prime}}$ is the Kronecker delta function. The left-hand side of equation \ref{lagragianformnewton} indicates the acceleration of body mass due to deformation and $F_V^{\prime}$ is the external force applied by the structure. 
The stress in the deformed structure was computed from the Cauchy stress tensor ($\sigma$) using the following formula \citep{hill1998mathematical}: 
\begin{equation}
    S = S_{add} + \textbf{C}:\epsilon_{el}. 
\label{Pkstress}
\end{equation}
In equation \ref{Pkstress}, $S_{add}$ indicates any extra stress rather than thermal stress, which can be initial, damping, and viscoelastic stresses. In our analysis, we considered that the proposed structure is initially in a stress-free state. Hence, the $S_{add}$ value is zero for our analysis. In equation \ref{Pkstress}, \textbf{C} represents the elasticity tensor and is contacted with another second-order tensor elastic strain ($\epsilon_{el}$). This portion ($\textbf{C}:\epsilon_{el}$) is called deviatoric stress and can be written as \citep{hill1998mathematical},
\begin{equation}
    \sigma_d = \textbf{C}(\textbf{E},\nu, \textbf{G}):\epsilon_{el}.
\label{eq8}
\end{equation}
Von Mises stress ($\sigma_{v}$) is related to the deviatoric stress ($\sigma_{d}$) of equation \ref{eq8} by, 
\begin{equation}
\sigma_v = \sqrt{\frac{3}{2} \sigma_{d}:\sigma_{d}}.
\label{eq9}
\end{equation}
The equation \ref{eq9} was used to compute the von Mises stress on the structure at different concentration factors of the incident radiation. The structure will permanently be cracked or fractured if the magnitude of the von Mises stress exceeds the ultimate tensile strength of the material.
In this work, the durability of the device was determined by analyzing the crack phase field, which measures the initiation and propagation of cracks within the structure (see Section S3 of Supplementary Material for more details). 
\\
\\


\section{Results and discussion}
\subsection{ Structural parameter optimization}
Various structural parameters like the number of metal-dielectric layers, layer width, and thickness were tuned to determine their optimum value at which the structure provided a broadband-perfect absorption. To maintain a symmetrical structure, the square-shaped layers were selected, and each upper layer was shortened by 20 nm from its bottom layer to form the pyramid. At first, the optimum number of layers was determined for an arbitrary width and metal-dielectric thickness. From Fig.\,\ref{FIG:2}(a), it can be seen that as the number of metal-dielectric layers increases, absorption increases. The absorption profile has a dip around 2500 nm wavelength, which reduces gradually with the increasing number of layers. Furthermore, a very small difference was found in the absorption profile for a total of 18 layers with t\textsubscript{M1} = 180 nm than 20 layers with t\textsubscript{M1} = 60 nm, as can be seen from Fig.\,\ref{FIG:2}(b). Besides, the topmost metal layer, with a thickness three times the other metal layers, provides high average absorption among the other configurations due to high absorption at longer wavelength regions. 

The metal-insulator-metal-insulator (MIMI) structure was chosen over the metal-insulator-metal (MIM) structure in this analysis because of the high average absorption of the MIMI structure at a shorter wavelength region, which can be observed in Fig.\,\ref{FIG:2}(c). Afterward, the optimum period of the structure was determined by changing the width of the bottom layer from 400 to 620 nm with a 10 nm interval. From Fig.\,\ref{FIG:2}(d), it can be observed that although the absorption profile has been extended to a larger wavelength range with the increasing period, the average absorption around 2500 nm decreases. The optimal period of the structure was chosen as 500 nm, considering the highest average absorption up to a wavelength of 5000 nm. To determine the optimum metal and dielectric thickness, the dielectric thickness was swept from 10 to 120 nm with a 2.5 nm interval, and the metal thickness was changed from 5 to 70 nm with a 2.5 nm interval. The absorption spectrum for each dielectric thickness was simulated for a certain metal thickness. The obtained absorption spectra have been shown in Figs.\,\ref{FIG:3}(a), (b), (c), and (d) for metal thicknesses 20, 40, 60, and 70 nm, respectively. It can be seen that the absorption at shorter wavelengths (200 -- 300 nm) is below 95\%, irrespective of the dielectric thickness. This indicates the weaker coupling of cavity mode and guided mode resonance (GMR) with the incident light. From the wavelength region 350 to 2500 nm, the absorption increased with the dielectric thickness. This is obvious since the effectiveness of GMR is related to the thickness. Considering the highest average absorption, the optimum dielectric and metal thicknesses were chosen as 17.5 and 60 nm, respectively.


\begin{figure*}
	\centering
		\includegraphics[scale=.38]{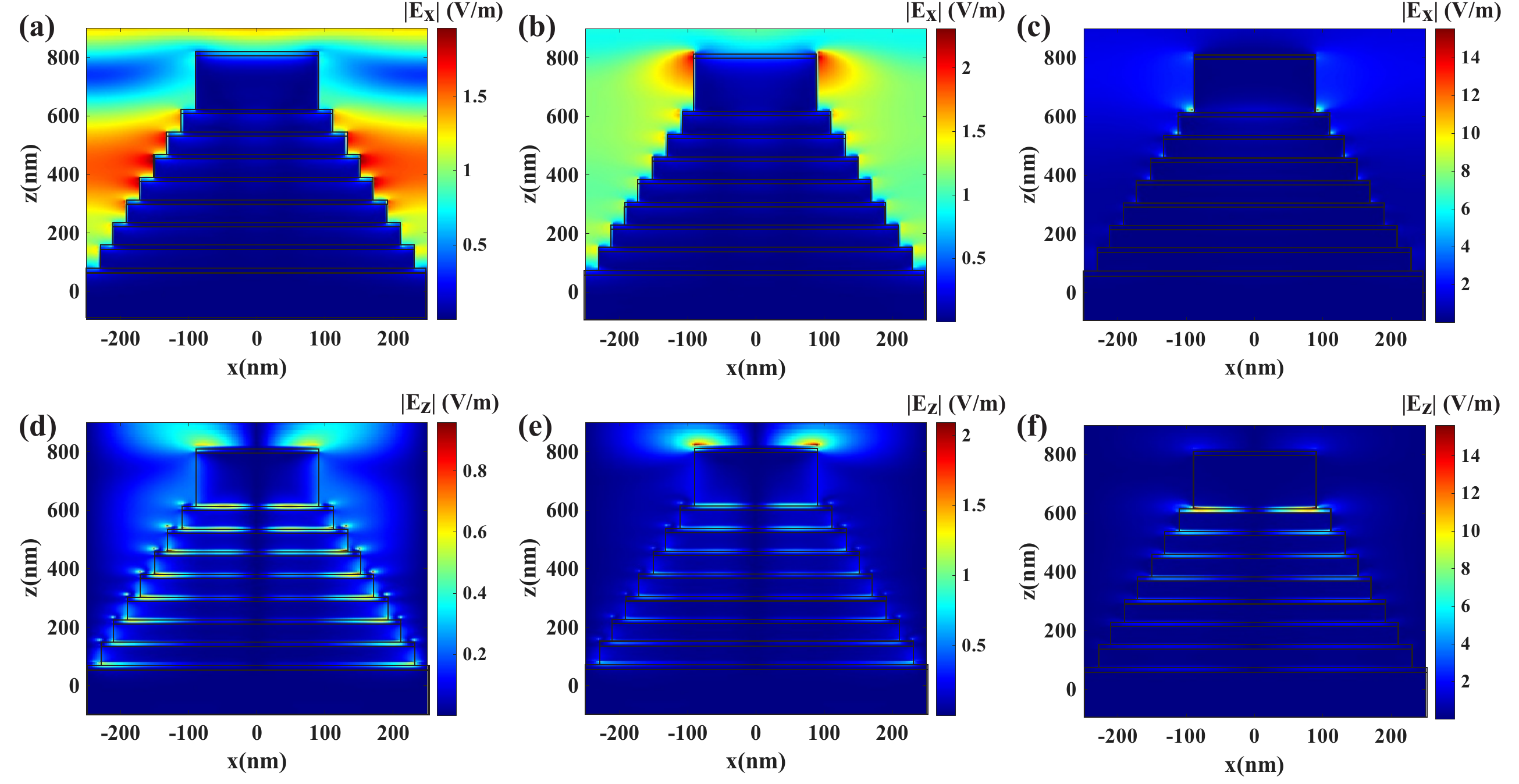}
	\caption{Electric field profile of the |Ex| component at (a) 490 nm, (b) 664 nm (c) 2230 nm wavelengths, and |Ez| component at (d) 490 nm, (e) 664 nm (f) 2230 nm wavelengths }
\label{FIG:7}
\end{figure*}
\subsection{Optical performance analysis}
\begin{figure*}
	\centering
		\includegraphics[scale=.38]{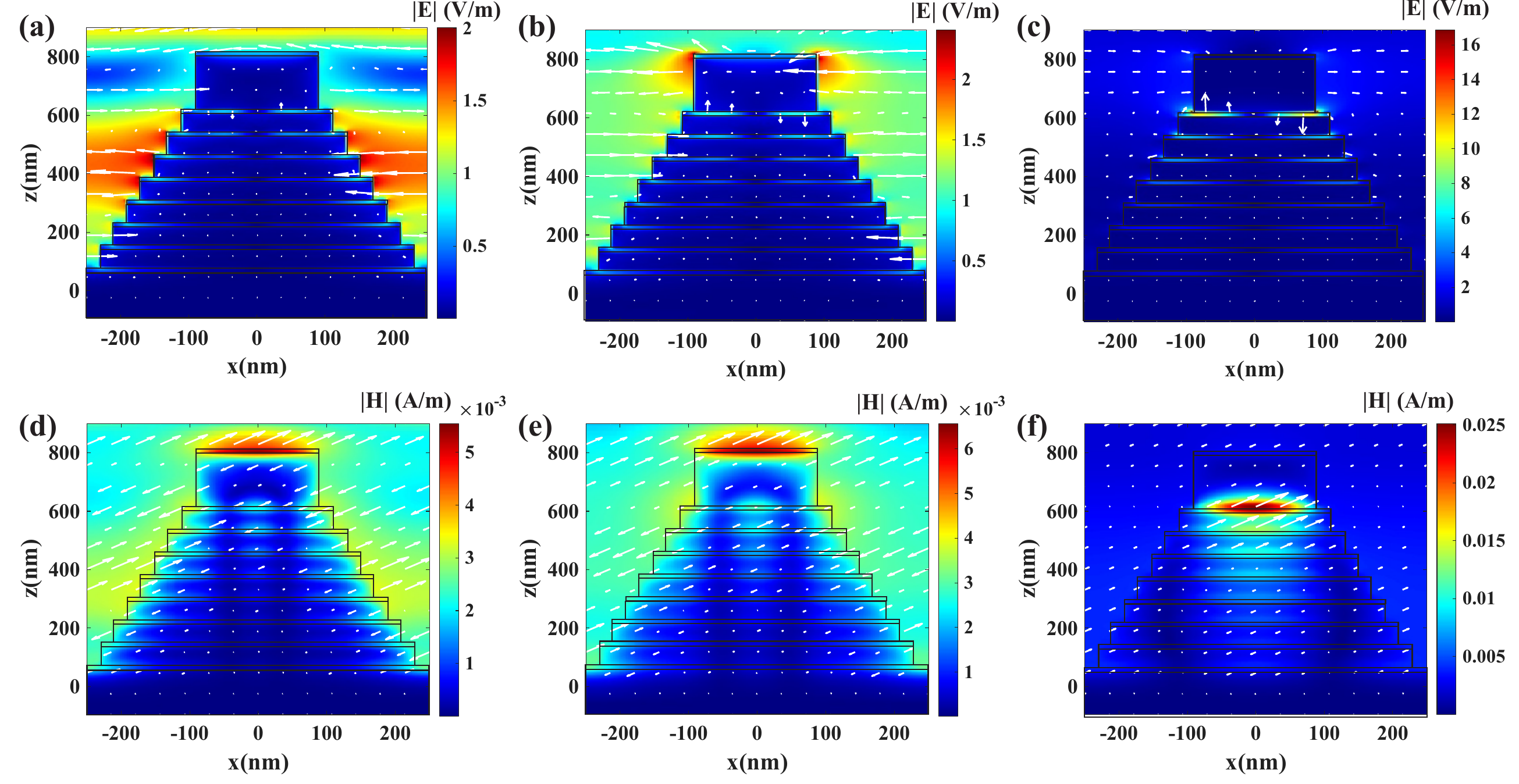}
	\caption{Total electric field profile with flux distribution at (a) 490 nm, (b) 664 nm, and (c) 2230 nm wavelengths, and total magnetic field profile with flux distribution at (d) 490 nm, (e) 664 nm, and (f) 2230 nm wavelengths.}
	\label{FIG:8}
\end{figure*}
\begin{figure*}[t]
	\centering
		\includegraphics[scale=.56]{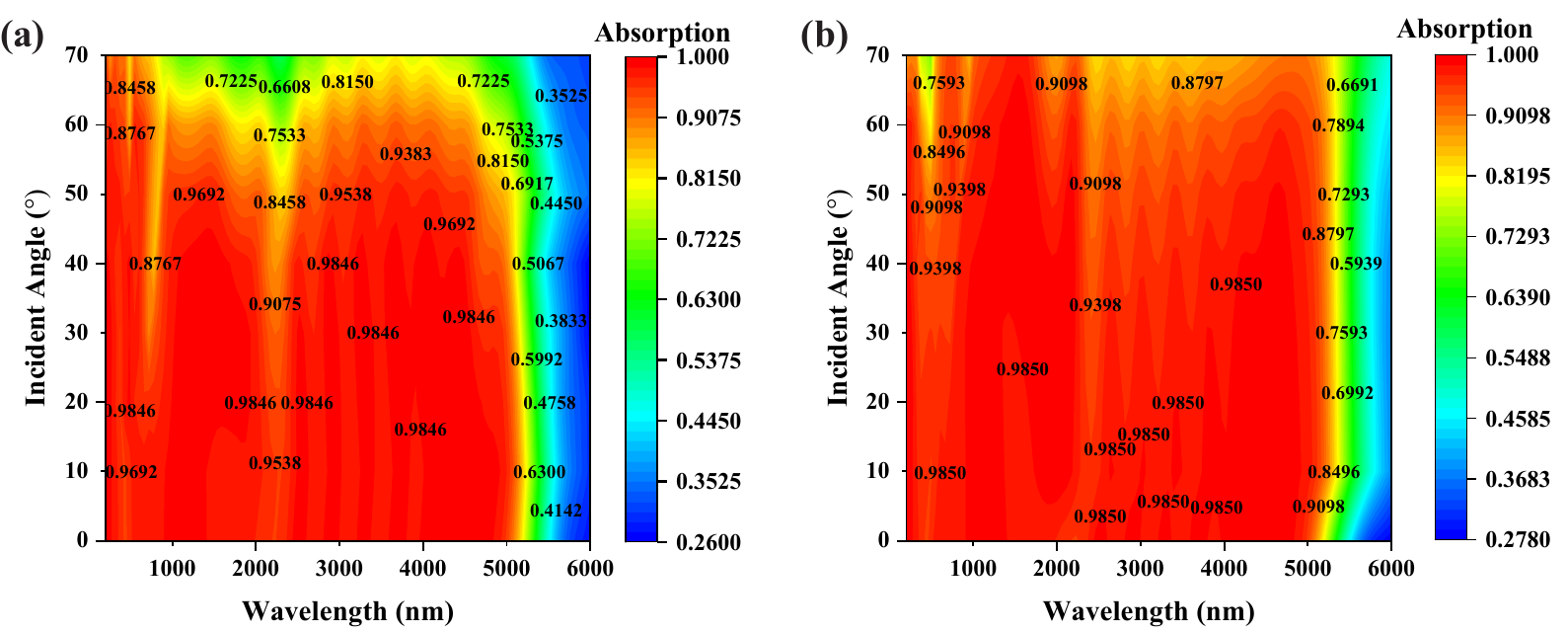}
	\caption{Effect of incident angle of the radiation on the absorption profile of the absorber for (a) TM and (b) TE polarized lights.}
\label{FIG:9}
\end{figure*}
\begin{figure*}
	\centering		\includegraphics[scale=.57]{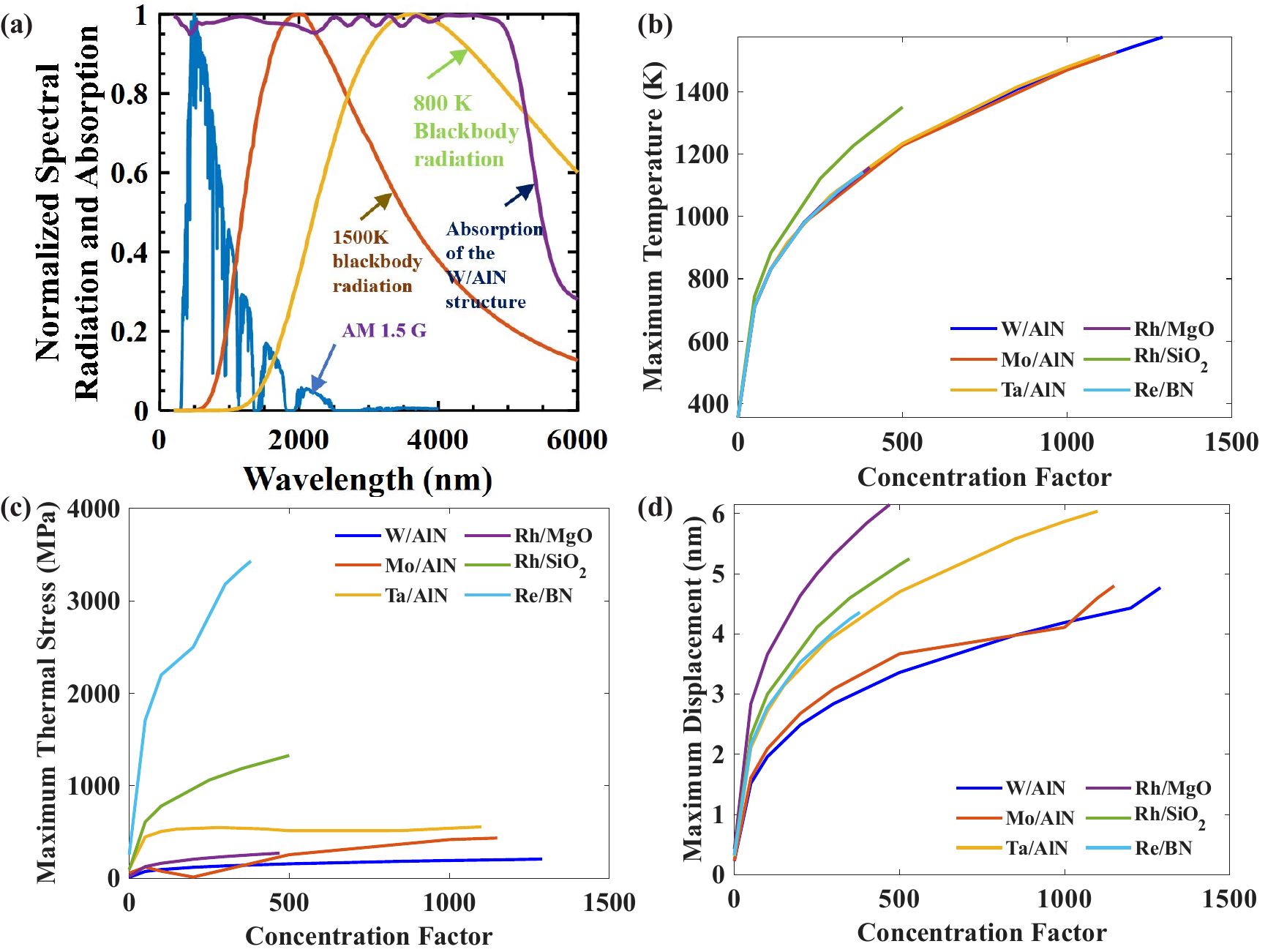}
	\caption{(a) Normalized spectral radiation from different sources, including the sun and other black body sources of various temperatures. Impact of different concentration factors on the maximum (b) temperature (c) thermal stress (d) displacement of the W/AlN, Mo/AlN, Ta/AlN, Rh/MgO, Rh/SiO\textsubscript{2}, Re/BN composite structures.}
\label{FIG:10}
\end{figure*}
The generation of broadband absorption is due to the individual contribution of multiple layers as well as due to the combination of multiple resonance modes at different frequency ranges. The contribution of individual metal and dielectric layers in the broadband absorption is illustrated in Figs.\,\ref{FIG:4} (a) and (b), respectively. From Figs.\,\ref{FIG:4} (a) and (b), it can be observed that the shorter wavelength light gets absorbed by the top layers, and the longer wavelength light is absorbed by the bottom layers because of the high penetration depth of the large wavelength light. Moreover, maximum absorption occurred at the metal layer than the dielectric, indicating that the dielectric layers mainly contributed to the SPR and decreased the speed of light to enhance the light-matter interaction time. Fig.\,\ref{FIG:5} depicts the absorption profile of different metal-dielectric structures with optimum dimensional parameters used in this study. From Fig.\,\ref{FIG:5}, it can be seen that the W/AlN, Mo/AlN, and Ta/AlN structures demonstrate over 90\% absorption for a longer wavelength while Rh/SiO\textsubscript{2}, Re/BN,and Rh/MgO structures provide over 85\% average absorption upto wavelength 3500 nm. Among all these different structures, W/AlN multilayer structure is capable of absorbing almost entire radiation spectrum upto wavelength 5000 nm. The average absorption of the used structures studied in this study is provided in Table \ref{tab:1}. From Table \ref{tab:1}, it can be seen that almost all the designed structures can absorb over 90\% of the incident radiation of the UV, visible, and near-IR regions. However, for all structures except W/AlN, average absorption drops below 90\% at the mid-IR region. The broadband absorption of the W/AlN structure will be further verified by determining the impedance matching, refractive index, permittivity, and permeability of the structure from scattering parameters and from the structure's electric and magnetic field profiles. 
A  good absorber must be impedance matched with the free space, equal to one. From Fig.\,\ref{FIG:6}(a), it can be seen that over 5000 nm wavelength, the real part ($\Omega^{\prime}$) and imaginary part ($\Omega$") of the effective impedance of the absorber is nearly one and zero, respectively, which indicates the impedance matching with free space. The positive imaginary part of the effective refractive index ($\kappa$) of the structure from Fig.\,\ref{FIG:6}(b) indicates the light through the structure is attenuating and absorbed throughout the wavelength range. The alternate positive and negative value of the real part of the refractive index (\textit{n}) indicates the refracted light's direction alternates as it propagates through the absorber, which indicates the possibility of cavity resonance modes up to approximately 2230 nm wavelength. The positive imaginary part of the effective permittivity and permeability of Figs.\,\ref{FIG:6}(c) and (d), respectively, refers to the high absorption and low reflection of the structure. The presence of both positive and negative real parts of the permeability ($\mu'$) and permittivity ($\epsilon'$) up to wavelength 2200 nm indicates that the incident light is absorbed not only through usual absorption modes but also through multiple reflections via cavity modes, which will be more clearly observed from the electric field analysis.
To evaluate the optical performance, the electric and magnetic fields of the proposed structure were analyzed for 490 nm, 664 nm, and 2230 nm, the onset of different resonance phenomena, for TM polarized light. From Figs.\,\ref{FIG:7}(a) and (d), it can be seen that high absorption at 490 nm is contributed by both \textbf{|E\textsubscript{x}|} and \textbf{|E\textsubscript{z}|} components. In this wavelength, the presence of \textbf{|E\textsubscript{x}|} at the corner of both sides of the nano-structure indicates the coupling of the electric field between the adjacent structures. Moreover, the presence of the electric field component \textbf{|E\textsubscript{z}|} within the subsequent dielectric layers indicated GMR and the Fabry-Perot (FP) cavity resonance, which is created by the two metal reflectors separated by the dielectric layer. At normal incidence, for the absorption peak at 664 nm, light was guided by the dielectric layers and forms a second-order GMR as described in Section S8 of the Supplementary Materials. At 490 nm, the value of the wavelength obeying the FP cavity resonance condition was approximately $495 \, \mathrm{nm}$, which was very close to the 490 nm wavelength and verified the existence of FP resonance within the structure at that wavelength. The detailed discussion of the origin of both of these references is provided in Section S8 of the Supplementary Material.

At 664 nm wavelength, the electric field is more concentrated towards the corners of the top metal layer, and the electric flux lines are circulated in the top metal layer as seen from Fig.\,\ref{FIG:7}(a). This is because of the localized surface plasmon resonance (LSPR). Same as the previous case, at this wavelength, \textbf{|E\textsubscript{x}|} is more prominent over \textbf{|E\textsubscript{z}|} in terms of field enhancement, which can be seen from Figs.\,\ref{FIG:7}(b) and (c). Fr\"ohlich resonance condition was used and provided in Section S8 of the Supplementary Materials to verify the existence of LSPR at 1186 nm wavelength, which was within the 664 to 2230 nm wavelength range. On the other hand, at 2230 nm wavelength, \textbf{|E\textsubscript{z}|} is more superior over \textbf{|E\textsubscript{x}|} as depicted in Figs.\,\ref{FIG:7}(c) and (f). Besides, \textbf{|E\textsubscript{z}|} component is responsible for the presence of the surface plasmon polariton (SPP) mode at this wavelength. From Fig.\,\ref{FIG:7}(c), it is seen that the E-field is concentrated within the top dielectric layer, and the electric flux lines are approaching both the metal and dielectric layer, which indicates the presence of the SPP mode. By determining the effective refractive index for SPP mode and comparing it with the refractive index of the dielectric material as described in Section S8 of the Supplementary Materials, the SPP coupling condition was verified at 2500 nm wavelength. From Figs.\,\ref{FIG:8}(d), (e), and (f), it can be seen that the magnetic field strength is almost 1000 times weaker compared to the electric field. Though the magnetic field is quite weaker, it also has a significant role in the absorption process. The H-field mainly controls the momentum of the incident wave. It can be observed that the H-field was highly concentrated within the dielectric layers because of its high refractive index at different wavelengths, which increased the magnetic coupling and greatly reduced the momentum of the incident wave and hence created a slow wave effect (SWE). Due to SWE, the electromagnetic wave interacted with the designed structure for a longer period, consequently improving absorption. Comparing Figs.\,\ref{FIG:8}(d) and (e), it can be concluded that the H-field is more concentrated within almost all dielectric layers at wavelength 490 nm. In contrast, it is mainly concentrated on the top and bottom dielectric layers of the top thick metal layer at wavelength 664 nm which verifies the presence of GMR, cavity resonance at 490 nm and the LSPR at wavelength 664 nm. Fig.\,\ref{FIG:8}(f) illustrates the confinement of light solely within the second dielectric layer and also the flux lines circulating it. These circulated flux lines provide a linear electric field at the top and bottom of this dielectric layer, which indicates the presence of SPP that can extend through the dielectric from the metal layer. In conclusion, the designed structure exhibits three different absorption phenomena at different wavelength regions. From 200 to 664 nm, the absorption is mainly due to GMR and FP resonance; from 664nm to 2230 nm, the absorption is solely on the top metal layer due to LSPR, and from 2230 nm onwards, the absorption occurs due to SPP at different metal-dielectric pairs.

The response of the proposed structure on different incident angles of the TM and TE polarized light has been illustrated in Figs.\,\ref{FIG:9}(a) and (b), respectively. For TM polarized light, the structure can absorb  >90\% of the incident light up to wavelength 5000 nm for 50$^\circ$ as depicted in \ref{FIG:9}(a). Above 50$^\circ$, the absorption of the incident spectrum decreased with increasing wavelength. On the other hand, as can be seen from \ref{FIG:9}(b), for TE polarized light, the average absorption increased slightly for low energy photons above 50$^\circ$ up to wavelength approximately 5000 nm, after that the absorption deceased.

\begin{table*}[width=2\linewidth,cols=4]
\caption{Ultimate tensile strength of different metal-dielectric composites at standard temperature.}\label{tbl1}
\begin{tabular*}{\tblwidth}{@{} LLL@{} }
\toprule
\makecell {Material} & \makecell{Ultimate \\Tensile Strength} & \makecell{Ref.}\\
\midrule
\makecell{Tungsten (W)} & \makecell{980 MPa} & \makecell{\citep{ross2013metallic}} \\
\makecell{Molybdenum (Mo)} & \makecell{1400 MPa} & \makecell{\citep{schauer2021biocompatibility}}\\ 
\makecell{Rhodium (Rh)} & \makecell{2068 MPa} & \makecell{\citep{ross2013metallic}}\\
\makecell{Rhenium (Re)} 
& \makecell{1070 MPa} & \makecell{\citep{ross2013metallic}}\\ 
\makecell{Aluminium Nitrite (AlN)} & \makecell{300 MPa} & \makecell{\citep{gang2004tensile}}\\
\makecell{Magnesium Oxide (MgO)} & \makecell{96 MPa} & \makecell{\citep{nobre2020magnesia}}\\
\makecell{Silicon Di-Oxide (SiO\textsubscript{2})} & \makecell{300 MPa-900 MPa} & \makecell{\citep{sharpe2001effect}}\\
\makecell{Boron Nitrite(BN)} & \makecell{41 MPa- 55 MPa} & \makecell{\citep{bauccio1994asm}}\\
\bottomrule
\label{tab:2}
\end{tabular*}
\end{table*}

\subsection{Thermal performance of the absorber}
At high temperature, the multilayer structure experiences large thermal expansion, but because the bonded layers have different coefficients of thermal expansion and are mechanically constrained, they cannot expand freely. This thermal strain is converted to stress via the elastic tensor according to equation \ref{Pkstress}.  These stresses require to be balanced across the structure according to equation \ref{lagragianformnewton} and this leads to a strong localization of stress near edges and interfaces. The effect of local field on the stress is defined by the Kirchhoff stress factor by equation \ref{eq5}. The deviatoric part of these   stresses drives the Von~Mises stress in accordance with equations \ref{eq8} and \ref{eq9}. At high temperature material's von~Mises stress exceeded ultimate tensile strength and certain regions of the structure experience permanent deformation where the elasticity could not be maintained.  
\begin{figure*}
	\centering
		\includegraphics[scale=.67]{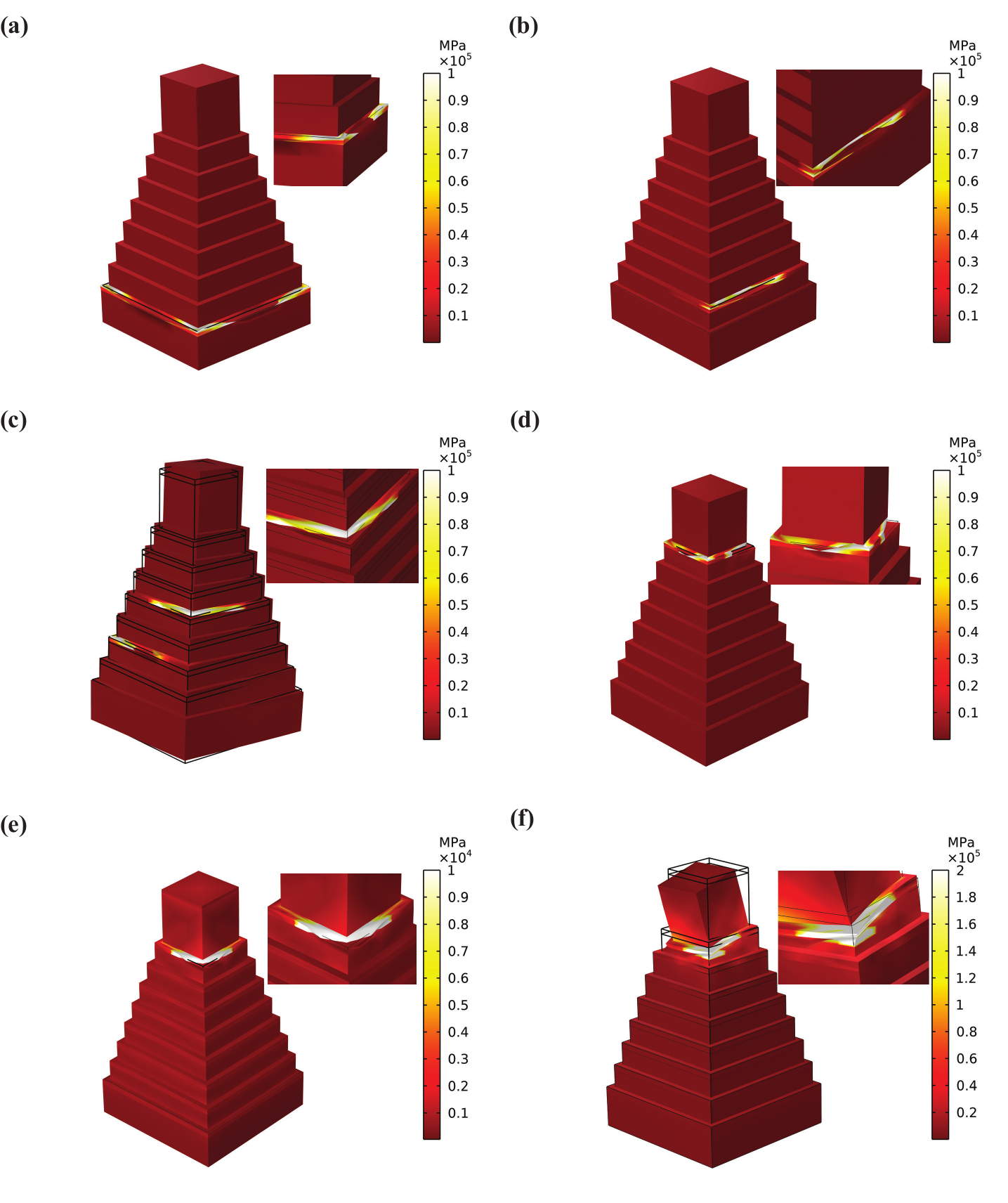}
	\caption{The permanent deformation of (a) W/AlN at 1500 concentration factor, (b) Mo/AlN at 1170 concentration factor, (c) Ta/AlN at 1150 concentration factor, (d) Rh/MgO at 650 concentration factor, (e) Rh/SiO\textsubscript{2} at  550 concentration factor, and (f) Re/BN at 390 concentration factor.}
\label{FIG:11}
\end{figure*}

We calculated the thermal performance of the proposed structure by determining the thermal stress on the structure. Since the efficiency of the TPV system is proportional to the temperature to a certain extent, the incident radiation was concentrated on the structure to improve the temperature and reduce the thermal loss. The structure will naturally exhibit thermal expansion at elevated temperatures, and the inner molecules of the structure may produce thermal stress against this expansion. 

Here, we determined the von Mises stress of the structure for different concentration factors and compared the stress obtained with the ultimate tensile strength of the structure. 
The ultimate tensile strength of the metal and dielectric materials used is provided in Table \ref{tab:2}. 
Fig.\,\ref{FIG:10} illustrates the maximum thermal stress, temperature, and displacement magnitude of different metal-dielectric composites at different concentration factors. 

Fig.\,\ref{FIG:10} illustrates the maximum thermal stress, temperature, and displacement magnitude of different metal-dielectric composites at different concentration factors. Fig.\,\ref{FIG:10} (a) illustrates the normalized spectral radiation of different black body sources of various temperatures along with the AM 1.5G radiation and indicates that the designed structure can absorb almost all of the incident radiation. From Fig.\,\ref{FIG:10} (b), it can be seen that maximum temperatures of all the composite structures follow the same trend, and even the magnitude of the maximum temperature is quite similar with changing concentration factor except for Rh/MgO, whose temperature is a few times higher than the others. This is because of the very low value of thermal conductivity of the SiO\textsubscript{2} as can be seen from Table S2 of Supplementary Material, which causes the heat to be localized and elevates the temperature at a faster rate. The change in the maximum thermal stress of the structure for changing the concentration factor is illustrated in Fig.\,\ref{FIG:10} (c). From Fig.\,\ref{FIG:10} (c), it can be seen that the thermal stress of Rh/SiO\textsubscript{2} and Re/BN increases rapidly with changing concentration factor, which leads to an early failure of these structures as can be seen from Figs.\,\ref{FIG:10} (e) and (f). This occurred due to the larger difference in the coefficient of thermal expansion between the two materials (see Table S2 of  Supplementary Material) of the respective composite structures, which produced higher stress at the interface between the layers. On the other hand, W/AlN, Mo/AlN, Ta/AlN, and Rh/MgO exhibited thermal stress below 1000 for a wide range of concentration factors because of closely matched thermal expansion coefficients, as can be observed from Table S2 of Supplementary Material. The effect of concentration factors on the displacement magnitude of each structure is demonstrated in Fig.\,\ref{FIG:10} (d). It can be observed that the maximum displacement from the original shape of the structure is always greater with increasing concentration factor. This happened due to the increasing concentration factor; temperature increased, which caused more thermal expansion (see Fig. S3 of the Supplementary Material for 3D-distribution of temperature, maximum con Mises stress, displacement magnitude, radiosity, and elastic energy density profile of the W/AlN structure at concentration factor 100).
The structure was subjected to permanent damage when the thermal stress exceeded the ultimate tensile strength of the material. Since the stress increases with temperature,  there must be a maximum concentration factor at which the composite structure will be cracked permanently. From Fig.\,\ref{FIG:11}(a),  it can be seen that at a concentration factor of 1500, the edge of the bottom AlN layers got fractured due to excessive thermal stress. This occurred since at high temperature, the multilayer structure experiences large thermal expansion, but because the bonded layers have different coefficients of thermal expansion and were mechanically constrained, they could not expand freely. This thermal strain was converted to stress via the elastic tensor according to equation \ref{Pkstress}.  These stresses required to be balanced across the structure according to equation \ref{lagragianformnewton}, and this led to a strong localization of stress near edges and interfaces. The deviatoric part of these stresses drives the Von~Mises stress in accordance with equations \ref{eq8} and \ref{eq9}. At high temperature, the material's von~Mises stress exceeded the ultimate tensile strength, and certain regions of the structure experienced permanent deformation where the elasticity could not be maintained. A similar type of defect occurs in the case of Mo/AlN composite structure at concentration factor 1170 as can be seen from Fig.\,\ref{FIG:11}(b). However, the first few layers of the Ta/AlN structure exhibited a sharp deviation from its initial position and defects at the edges of the middle AlN layers at a concentration factor of 1150, which can be observed from Fig.\,\ref{FIG:11}(c). This large deformation can be due to the large local stress, which can be defined by the Kirchhoff stress factor of \ref{eq5} and causes additional ruptures of the structure. Fig.\,\ref{FIG:11}(d) illustrates that the second MgO layer gets completely damaged from all ends in the Rh/MgO composite structure for concentration factor 650. A similar damage pattern at the second SiO\textsubscript{2} layer occurred for Rh/SiO\textsubscript{2} structure at concentration factor 550 due to an excessive compressive stress, as can be seen from Fig.\,\ref{FIG:11}(e). The Re/BN  performed poorly among all the proposed structures. From Fig.\,\ref{FIG:11}(f), it can be observed that at concentration factor 390, both second layers of Re/BN completely damaged at a certain end, which caused the structure to lean over that end.  
Table \ref{tab:3} demonstrates the maximum permissible concentration factor and maximum thermal stress, temperature, and displacement magnitude of the composite structure at that concentration factor. It can be observed from Table \ref{tab:3} that the W/AlN can withstand the highest concentration factor as well as temperature among all structures. On the contrary, Re/BN performed the worst. Besides, the structures with AlN as the dielectric performed better due to the alignment of the thermal expansion coefficient of the AlN with W, Ta, and Mo compared to other composite structures. Besides from the detailed analysis based on the metal-oxidation as mentioned in Section S5 of the Supplementary Materials, it can be concluded that W and Rh can be stable at high vacuum, low-pressure conditions at a very high temperature.
\begin{table*}[width=2\linewidth,cols=4]
\caption{Thermomechanical properties of different structures at maximum allowable concentration factor.}\label{tbl1}
\begin{tabular*}{\tblwidth}{@{} LLLLL@{} }
\toprule
\makecell {Structure} & \makecell{Maximum Allowable \\Concentration Factor} & \makecell{Maximum \\ Temperature\\ (K)} & \makecell{Maximum \\ Thermal Stress \\(MPa)} & \makecell{Maximum \\ Displacement \\ (nm)}\\
\midrule
\makecell{W/AlN} & \makecell{1490--1500} & \makecell{1687-- 1704} & \makecell{223--225} & \makecell{4.71-- 4.75}\\
\makecell{Mo/AlN} & \makecell{1160--1170} & \makecell{1533--1535} & \makecell{435--437} & \makecell{4.81}\\ 
\makecell{Ta/AlN} & \makecell{1140--1150} & \makecell{1534--1538} & \makecell{554--560} & \makecell{6.02--6.1}\\
\makecell{Rh/MgO} 
& \makecell{640--650} & \makecell{1277--1280} 
& \makecell{294--300} & \makecell{6.61-- 6.65}\\ 
\makecell{Rh/SiO\textsubscript{2}} & \makecell{540-550} & \makecell{1371--1375} & \makecell{1352--1360} &\makecell{5.11-- 5.3} \\
\makecell{Re/BN} & \makecell{390--400} & \makecell{1137-- 1145} & \makecell{3400--3430} &\makecell{4.3-- 4.36} \\
\bottomrule
\end{tabular*}
\label{tab:3}
\end{table*}
\begin{figure*}
	\centering
		\includegraphics[scale=.57]{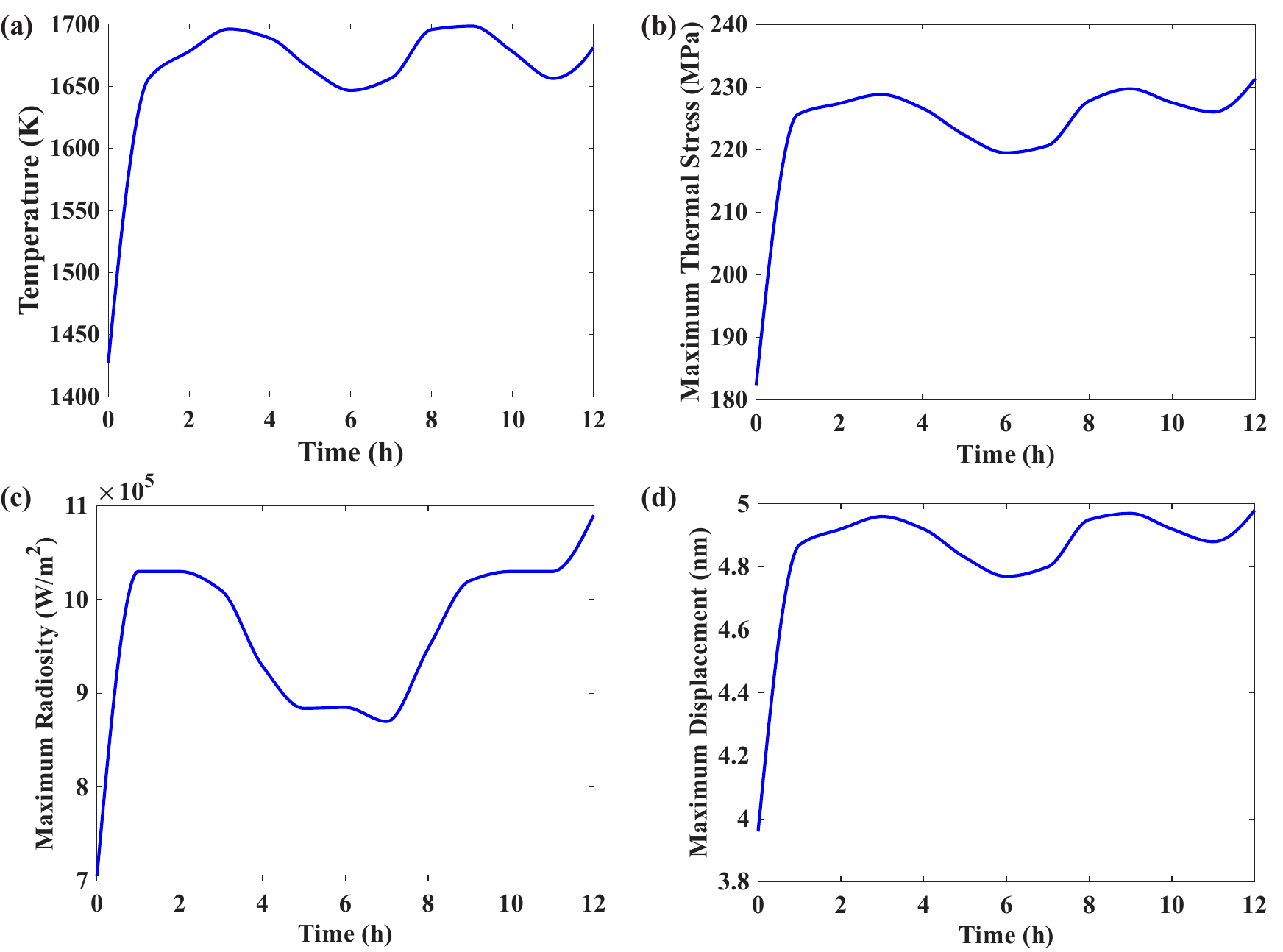}
	\caption{The effect of continuous thermal loading of 1490 concentration factor on the W/AlN composite structure's maximum (a) temperature (b) thermal stress(c) radiosity (d) displacement magnitude. The inset figure of (a) illustrates the spectral emission of the proposed absorber is spectrally matched with the 1700 K blackbody radiation.}
\label{FIG:12}
\end{figure*}

A time-dependent analysis was performed on the W/AlN structure to determine the effect of constant thermal loading over an extended period of 12 h. From Fig.\ref{FIG:12}(a), it can be observed that the structure temperature did not reach a steady-state value; rather, it fluctuated over a period of time. This can be due to the low and high heat capacity of W and AlN, respectively, as can be seen in Table S2 of the Supplementary Material. The periodic fluctuation of temperature causes periodic expansion and contraction, and this provides an increase and decrease in the magnitude of the thermal stress, as can be seen in Fig.\ref{FIG:12}(b). This indicates the elastic behaviour of the structure, as can be observed in Fig. \ref{FIG:12} (d). Since the radiated energy from the heated body is proportional to the fourth power of the temperature, the radiosity of the designed absorber follows its temperature profile over the time period as illustrated in Fig. \ref{FIG:12}(c).

\subsection{Effect of temperature on the optical performance}
From optical and thermal analysis, we observed that among all structures, the W/AlN structure exhibited high average absorption and high thermal stability. For this reason, the temperature-dependent optical performance analysis in this section is limited to the W/AlN structure only. To determine the optical performance of the absorber at high temperature, a temperature-dependent Drude-Lorentz model was used to evaluate the complex relative permittivity and eventually the complex refractive index as follows \citep{silva2025numerical},
\begin{equation*}
\varepsilon_{r}(T) = 1 - \frac{f_{0}\omega_{p}^{2}(T)}{\omega\big(\omega - i\Gamma_{D}(T)\big)} 
+ \sum_{j=1}^{N} \frac{f_{j}\omega_{p}^{2}(T)}{\omega_{j}^{2}(T) - \omega^{2} - i\omega\Gamma_{j}(T)}. \tag{11}
\end{equation*}
Here, $\omega_{p}$ and $\omega_{f}$ are the plasma frequency and oscillator's frequency, respectively; N is the number of oscillators; $\Gamma_f$ and $\Gamma_j$ are the damping constants and $f_0$, $f_j$ are the oscillator's strength. The value of different Drude-Lorentz parameters,  $\omega_j$, $f_j$, and $\Gamma_j$ for W is provided in the Table S3 of the Supplementary Material.

\begin{figure*}
	\centering		\includegraphics[scale=1]{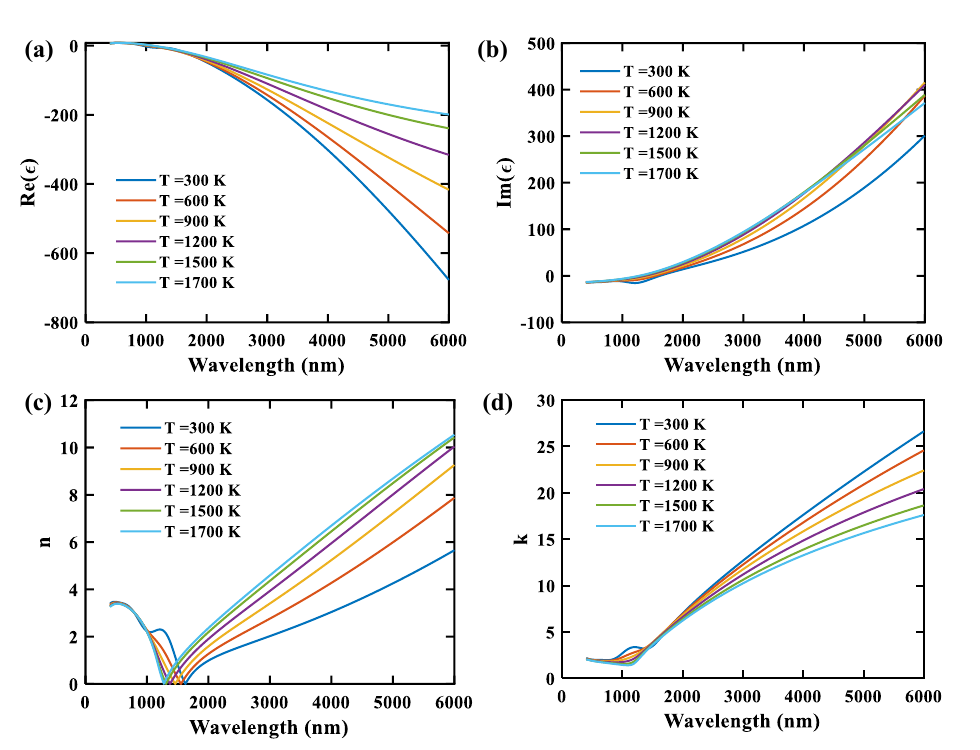}
	\caption{Temperature-dependent (a) real part of complex permittivity, (b) imaginary part of complex permittivity, (c) refractive index, and (d) extinction coefficient of tungsten.}
\label{FIG:13}
\end{figure*}
\begin{figure*}
	\centering		\includegraphics[scale=1]{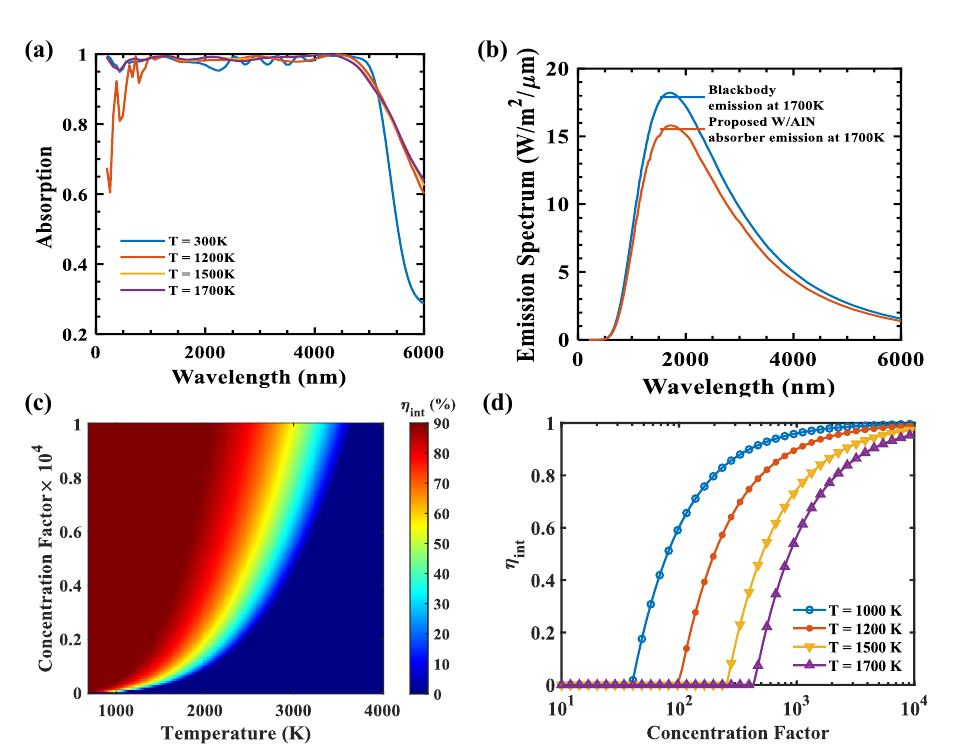}
	\caption{(a) Absorption spectra of W/AlN structure at various temperatures. (b) Comparing the emission spectrum of the proposed W/alN absorber at 1700K with the blackbody radiation, (c) Intermediate
efficiency of the W/AlN absorber ($\eta_{int}$) as a function of solar concentration and the temperature of nanosheet structure. (d) $\eta_{int}$ at different temperatures with the solar concentration at a logarithmic scale.}
\label{FIG:14}
\end{figure*}
\noindent
The temperature-dependent complex relative permittivity was determined through several steps. Firstly, the damping coefficient of the Drude oscillator due to electron-phonon and electron-electron interactions was determined using the following formula \citep{silva2025numerical}:
\begin{equation}
    \Gamma_{D}(T) = \Gamma_{e-e}(T) + \Gamma_{e-ph}(T)
\end{equation}
In the above equation, $\Gamma_{e-ph}(T) = \Gamma_{0}\!\left[\tfrac{1}{2} + \tfrac{T}{T_{D}}\right]$ and $\Gamma_{e-e}(T) = \dfrac{\pi^{3}}{12\hbar E_{f}}\big[(k_{B}T)^{2}+(\hbar\omega)^{2}\big]$, 
where $E_{f}$ is the Fermi energy, $\hbar$ is the reduced Planck’s constant, and 
$T_{D}$ is the Drude temperature. The effect of temperature on the plasma frequency of the Drude oscillator according to the thermal expansion coefficient $(\gamma)$ was determined by \citep{silva2025numerical},
\begin{equation}
    \omega_{p}(T) = \dfrac{\omega_{p}}{\sqrt{1+\gamma(T-T_{\text{r}})}}
\end{equation}
Here, $T_r$ is the room temperature. Lastly, the Lorentz oscillator's damping coefficient and the resonant frequencies were determined utilizing the following equations \citep{silva2025numerical}.
\begin{equation}
    \Gamma_{j}(T) = \Gamma_{j} + \alpha(\sqrt{T}-\sqrt{T_{\text{r}}}),
\end{equation}
\begin{equation}
    \omega_{j}^{2}(T) = \omega_{j}^{2} - \beta(\sqrt{T}-\sqrt{T_{\text{r}}})
\end{equation}
The temperature-dependent parameters of the Drude-Lorentz oscillator are provided in Table S4 of the Supplementary Material.
After obtaining the real (Re$(\epsilon)$) and imaginary (Im$(\epsilon)$)part of complex permittivity, the refractive index (n) and extinction coefficient (k) were calculated as follows \citep{Fox}: 
\begin{equation}
    n = \sqrt{\frac{\sqrt{\big(\mathrm{Re}(\varepsilon)\big)^{2} + \big(\mathrm{Im}(\varepsilon)\big)^{2}} + \mathrm{Re}(\varepsilon)}{2}}
    \label{eqn}
\end{equation}
\begin{equation}
   k = \sqrt{\frac{\sqrt{\big(\mathrm{Re}(\varepsilon)\big)^{2} + \big(\mathrm{Im}(\varepsilon)\big)^{2}} - \mathrm{Re}(\varepsilon)}{2}}
    \label{eqk}
\end{equation}
\\
High temperature significantly affects the complex permittivity and the emittance of the W. Figs.\, \ref{FIG:13} (a) and (b) demonstrate the real and imaginary parts of the complex permittivity of tungsten, respectively, for various temperatures. At wavelengths above 1000 nm, both the real and imaginary parts of the permittivity of tungsten, generated using the Drude-Lorentz model, match the extracted values obtained from the Palik handbook at a 300 K temperature. However, for longer wavelengths, the model exhibits differences depending on the temperature. From Fig.\,\ref{FIG:13} (a), it can be seen that as the temperature increases, Re$(\epsilon)$ decreases. On the contrary, from Fig.\,\ref{FIG:13} (b), it can be observed that the value of Im$(\epsilon)$ increases with increasing temperature. Due to such temperature dependent characteristics of Re$(\epsilon)$ and Im$(\epsilon)$ and according to equations \ref{eqn} and \ref{eqk}, both refractive index and extinction coefficients of tungsten increase at longer wavelengths with increasing temperature as can be seen from Figs.\,\ref{FIG:13} (c) and (d), respectively. This temperature-dependent behaviour can be attributed to the fact that, at higher temperatures, electron-phonon interactions increase, contributing to intraband transitions (Drude model) in the infrared region. However, this effect is minor in interband transitions (Lorentz model) in the visible range. Due to such increasing value of refractive index and extinction coefficient with temperature, the light heavily attenuates within the structure and contributes to high absorption/emission at high temperatures, as can be observed from Fig.\,\ref{FIG:14} (a). Fig.\,\ref{FIG:14} (b) represents the the power loss due to thermal emission of our proposed W/AlN absorber by matching the spectral emission ($I(\lambda, T)$) of the 1700K heated absorber with the blackbody emission ($I_{\mathrm{BB}}(\lambda, T)$) by using the following formula: 
\begin{equation}
    I(\lambda, T) = \varepsilon_a(\lambda) \cdot I_{\mathrm{BB}}(\lambda, T)
\end{equation}
Where, $\varepsilon(\lambda)$ is the emittance of the W/AlN structure,  $\lambda$ is the wavelength, and
$I_{\mathrm{BB}}$ represents the heated absorber radiation at equilibrium temperature $T$,. The overall TPV system's efficiency depends on two important factors. First one is the
intermediate efficiency ($\eta_{int}$) considering the effect of
absorber self-emission loss as a function of structure temperature
and concentration factor and the second performance factor is the effciency of the solar cell ($\eta_{PV}$). The $\eta_{int}$ was determined using the following
equation \citep{cui2024highly}:
\begin{equation}
\eta_{\mathrm{int}} =
\frac{P_{\mathrm{s}} - P_{\mathrm{a}}}{P_{\mathrm{s}}}
= 1 - \frac{P_{\mathrm{a}}}{P_{\mathrm{s}}}
\label{eq:018}
\end{equation}
Where, $P_{\mathrm{s}}$ and $P_{\mathrm{a}}$ are the incident powers from the blackbody or solar source and the emitted self-irradiance from the heated absorber, respectively. The expressions for the $P_{\mathrm{s}}$ and $P_{\mathrm{a}}$ are as follow \citep{cui2024highly}:

\begin{equation}
\begin{split}
P_{\mathrm{s}}(N) = &
\int_{0}^{2\pi} d\phi \int_{0}^{\theta_c} \sin\theta \cos\theta \, d\theta \\
& \times \int_{0}^{\infty} \varepsilon_{\mathrm{a}}(\lambda,\theta,\phi)\,
I_{\mathrm{s(BB/AM1.5)}}(\lambda,T)\, d\lambda
\end{split}
\label{eq:Psol}
\end{equation}

\begin{equation}
\begin{split}
P_{\mathrm{a}}(T) =\,
& \int_{0}^{2\pi} d\phi 
  \int_{0}^{\pi/2} \sin\theta \cos\theta \, d\theta \\
& \times \int_{0}^{\infty} 
  \varepsilon_{\mathrm{a}}(\lambda,\theta,\phi)\,
  I_{\mathrm{BB}}(\lambda,T)\, d\lambda
\end{split}
\label{eq:Pabs}
\end{equation}
Where $\theta$ and $\phi$ are the angles of incidence and polarization,
$\theta_c = \sin^{-1}\!\sqrt{\tfrac{N \Omega_s}{\pi}}$ where $N$ is the
concentration factor and $\Omega_s = 68.5~\mu\mathrm{sr}$ is the solid angle subtended by
the sun, 
and $I_{\mathrm{s}}$ is the radiation spectrum of the source that contains AM1.5G solar radiation along with the radiation of blackbody heat source. 
From Fig.\,\ref{FIG:14} (c), it can be seen that $\eta_{int}$ decreases
with increasing operating temperature at a steady  concentration factor and after a certain temperature, no efficiency is obtained. This is because of the fact that at a steady concentration factor, the incident radiation power from the source is constant but the emission loss increases as the fourth power of increasing temperature according to the Stefan-Boltzmann law. So, the effciency drops as the temperature increases according to the equation \ref{eq:018}. But for increasing concentration factor $\eta_{int}$ increases as the input power increases compared to the emission loss for a certain temperature, as can be seen from Fig.\,\ref{FIG:14} (d). But at low concentration factor, the input power from the source is low compared to the thermal emission loss which contributes to very low $\eta_{int}$. According to the Fig.\,\ref{FIG:14}, the emitter is in thermal contact with the absorber. Thus the emitter is generally heated to the same equilibrium temperature as the absorber. Considering the absorber's radiating surface temperature at 1700 K, the peak spectral wavelength from Wien's law is as follows \citep{chen2024review}:
\begin{equation}
    \lambda_{peak}\times T = 2898 (\mu m.K). \\
    \label{wl}
\end{equation}
According to equation \ref{wl} and from Fig.\,\ref{FIG:14} (b), it can be concluded that the emitter connected to our designed W/AlN structure can emit the maximum amount of photons at 1704 nm wavelength while radiated at 1700K temperature. This peak wavelength is perfectly matched with the solar cell material GaSb, having a bandgap of 0.73 eV. 
The optical efficiency of the proposed W/AlN absorber ($\eta_{absorber}$) at a certain temperature 
was determined by taking the ratio of the absorption multiplied 
by the incident irradiance to the total spectral radiation, as follows \citep{rana2021revisiting}:

\begin{equation}
\eta_{absorber} = \frac{\int_{0}^{\infty} \varepsilon_{a}(\lambda,\theta,\varphi) I_{sol(BB/AM1.5)}(\lambda,T)\, d\lambda}
{\int_{0}^{\infty} I_{sol(BB/AM1.5)}(\lambda,T)\, d\lambda}
\end{equation}

The $\eta_{absorber}$ was thereafter calculated as 98.68\% for  AM 1.5 spectrum (up to 4000 nm wavelength) and 97.73\% for the blackbody radiation at a certain temperature (up to 5072 nm wavelength) for normal incidence, respectively.
\begin{table*}[width=2\linewidth,cols=4]
\caption{Performance comparison of previous works with the proposed model.}\label{tbl1}
\begin{tabular*}{\tblwidth}{@{} LLLLL@{} }
\toprule
\makecell {Structure} & \makecell{Operating \\Wavelength} & \makecell{A\textsubscript{avg}} & \makecell{Incident\\Angle Tolerance} & \makecell{Ref.}\\
\midrule
\makecell{Ti/SiO\textsubscript{2} based \\ 2D-square bilayer\\ grating absorber} & \makecell{300nm-\\2100nm} & \makecell{98.3\%} & \makecell{-} & \makecell{\citep{li2023theoretical}}\\
\makecell{Ti/SiO\textsubscript{2} \\ nanoarrays} 
& \makecell{300nm-\\3000nm} & \makecell{99.87\%} 
& \makecell{-} & \makecell{\citep{qin2022broadband}}\\ 
\makecell{W/SiO\textsubscript{2} based \\multilayer absorber} & \makecell{500nm-\\3539nm} & \makecell{90\%} & \makecell{50$^\circ$} & \makecell{\citep{li2019ultra}}\\ 
\makecell{Si/Fe based \\rectangular shaped \\multilayer absorber} & \makecell{300nm-\\3000nm} & \makecell{96\%} & \makecell{70$^\circ$} & \makecell{\citep{liu2020ultra}}\\
\makecell{Cr-based \\ultrathin  square supercell-shaped\\ meta-absorbers} & \makecell{upto 1425 nm} & \makecell{>80\%} & \makecell{-} &\makecell{\citep{Ijaz2025}} \\
\makecell{W/HfO\textsubscript{2} based grating structure} & \makecell{300-2400 nm} & \makecell{>94\%} & \makecell{-} &\makecell{\citep{Feyisa2025}} \\
\makecell{Tapered pyramid\\ shape W/AlN composite structure} & \makecell{200nm - 5072 nm} & \makecell{97.73\%} & \makecell{60$^\circ$} &\makecell{This work} \\
\bottomrule
\end{tabular*}
\label{tab:4}
\end{table*}

\subsection{Comparative analysis}
The simultaneous utilization of both industrial waste heat and solar energy using a TPV system has not yet been reported; rather, most of them focused on the solar thermophotovoltaic (STPV) system. In Table \ref{tab:4}, various performance parameters of the previously studied STPV systems with broadband absorbers are highlighted compared to our proposed design. SiO\textsubscript{2}/Ti-based 2D grating structure proposed by Cai \textit{et al.} showed an average absorption of 98.3\% within the wavelength of 300 nm to 2100 nm \citep{li2023theoretical}. The W/SiO\textsubscript{2} IMI structure over the MIM structure was designed to be operated with 90\% absorption from 500 nm to 3539 nm wavelength and over 80\% absorption up to an incident angle 50$^\circ$ \citep{li2019ultra}. Liu \textit{et al.} reported a Si/Fe-based multilayer wideband absorber with an average absorptivity of 96\% up to a wavelength of 3000 nm \citep{liu2020ultra}. In 2025, Ijaz \textit{et al.} proposed a uniquely designed Cr-based ultrathin square-shaped meta absorber with a poor average absorption of around 80\% up to 1425 nm wavelength \citep{Ijaz2025}. Recently, a W/HfO\textsubscript{2} grating structure-based absorber with a broad spectral range up to 2400 nm with an average absorption of 94\% was proposed by Feyisa \textit{et al.} \citep{Feyisa2025}.
However, most of these aforementioned structures can not be used to utilize the waste heat generated from various industries along with solar energy because of the low range of temperatures (800--2500 K) of waste heat compared to the sun (around 5760 K), which leads to the generation of peak radiation from industrial waste heat at longer wavelengths, generally above 3000 nm. On the contrary, our proposed tapered-shape pyramid structure was capable of effectively absorbing 97.73\% of the radiation up to wavelength 5072 nm for the W/AlN structure. Besides, the proposed structures can provide over 96\% average absorption up to 50$^\circ$ incident angle and an average absorption of 85\%  up to an incident angle of 60$^\circ$ for both TM and TE polarized light. Moreover, compared to most of the other previous studies, our designed structure is polarization-independent and capable of efficiently producing almost similar responses to the polarized light at unpolarized conditions, at which the radiation spectrum is released from sources.

\subsection{Proposed Fabrication Technique}
The pyramid-shaped W/AlN multilayer emitter with 18 layers in total can be fabricated by adapting sputtering methods demonstrated for W/HfO\textsubscript{2} metamaterial stacks~\citep{chirumamilla2019metamaterial} and AlN thin-film deposition in \citep{kumagai2005growth}. A 120 nm thick tungsten layer can be fabricated on a sapphire substrate using DC magnetron sputtering in Ar at a very low base pressure to avoid oxidation. On the other hand, an AlN layer of thickness can be grown over the sapphire substrate by hydride vapor-phase epitaxy (HVPE) using AlCl\textsubscript{3} and NH\textsubscript{3} gases as precursors while maintaining a growth temperature ranging from 950 $^\circ$C to 1100 $^\circ$C to improve the crystalline quality. Above the bottom W layer, subsequent layers of AlN nm and W can be deposited by using plasma-enhanced atomic layer deposition (ALD)  or RF reactive sputtering in Ar/N\textsubscript{2} and DC sputtering, respectively. To form the staircase-shaped-pyramid structure, each bilayer of W and AlN can be patterned by using EBL-defined or PECVD hard masks . AlN can be etched by Ar/Cl\textsubscript{2}/BCl\textsubscript{3} ICP-RIE \citep{rammal2019aln}, and SF\textsubscript{6}/O\textsubscript{2} RIE can be applied to etch the W layer \citep{peignon1991etching}. The mask needs to be stripped, and an in situ Ar cleaning can be used to clean the surface before another redeposition. This cycle needs to be repeated until the desired number of W and AlN cap layers are deposited as a staircase-like architecture.  After each deposition, a post-annealing must be performed at 1000-1200 $^\circ$C under high vacuum pressure to stabilize W and densify AlN. Similarly, other proposed metamaterial structures can be fabricated.
\section{Conclusion}
We proposed and comprehensively analyzed highly durable, efficient, and broadband metamaterial absorbers for utilizing the industrial waste heat along with the solar energy using a thermophotovoltaic system. Depending on the high-temperature stability, thermomechanical properties, and lattice parameters, W/AlN, Ta/AlN, Mo/AlN, Rh/SiO\textsubscript{2}, Rh/MgO, and Re/BN metal-dielectric structures were chosen for this study. Among them, the proposed tapered-pyramid-shaped metamaterial absorber composed of W/AlN outperforms other composites in terms of optical performance and thermal stability. The designed W/AlN absorber was insensitive to polarization, absorbed around 97.73\% of the incident light up to 5072 nm wavelength, and maintained an average absorption of more than 85\% for up to 60$^\circ$ incident angle for both TM and TE polarized light, which is an encouraging attribute for the proposed metamaterial absorber. From the temperature-dependent Drude-Lorentz analysis,  we found that with an increase in the operating temperature, both the refractive index and extinction coefficient of W increased, and thus the absorption also increased, indicating high absorption/emission at high temperature. Apart from exciting optical performance, the designed structure exhibits better thermal stability over a high range of elevated temperatures. The crack phase field study reveals that before significant structural damage, the W/AlN absorber can withstand nearly 1500 times the incident radiation, and at that point, the structure's temperature will be around 1700 K. Furthermore, thermal analysis revealed that achieving high internal efficiency of the TPV system at elevated operating temperatures requires larger concentration factors for large input power, as the absorber experiences increased emission losses with rising temperature.
Our proposed ingenious design of pyramid-shaped multilayer W/AlN metamaterial absorber and the insights from the comprehensive performance analysis can be beneficial for the creation of an efficient component for next-generation thermophotovoltaic systems, capable of sustaining extreme operating temperatures. In addition to energy harvesting, the proposed absorber has enormous potential in other high-temperature applications like sensing, hypersonic, and aerospace applications.
\printcredits
\section*{Data Availability Statement}
{The data supporting the findings presented in this paper are not currently available to the public, but they may be obtained from the authors upon reasonable request.} 

\section*{Acknowledgements}
The authors thank the Bangladesh University of Engineering and Technology (BUET) for providing technical support. 

\bibliographystyle{unsrt}
\balance
\bibliography{main}

\end{document}